\documentclass{svjour3}                     
\smartqed  
\usepackage{graphicx}
\usepackage[top=30truemm,bottom=30truemm,left=25truemm,right=25truemm]{geometry}
\usepackage{amsmath}
\usepackage{amsfonts}
\usepackage{bm}
\usepackage{color}

\definecolor{blue}{rgb}{0, 0.4470, 0.7410}
\definecolor{red}{rgb}{0.8500, 0.1250, 0.0480} 
\definecolor{green}{rgb}{0.4660, 0.6740, 0.1880}
\graphicspath{{./Figures/}}

\journalname{Neural Comput. Appl.}
\begin{document}

\title{Generalization techniques of neural networks
for fluid flow estimation
}


\author{Masaki Morimoto \and
        Kai Fukami \and\\
        Kai Zhang \and
        Koji Fukagata 
}


\institute{M.Morimoto, K. Fukagata \at
              Department of Mechanical Engineering, Keio University,  Yokohama, 223-8522, Japan \\
              Tel.: +81-90-2240-6705\\
              \email{masaki.morimoto@kflab.jp}           
           \and
           K. Fukami \at
              Department of Mechanical and Aerospace Engineering, University of California, Los Angeles, CA 90095, USA\\
              Department of Mechanical Engineering, Keio University,  Yokohama, 223-8522, Japan
           \and
           K. Zhang \at
           Department of Mechanical and Aerospace Engineering, Rutgers University, Piscataway, NJ 08854, USA
}

\date{Received: date / Accepted: date}

\maketitle

\begin{abstract}
We demonstrate several techniques to encourage practical uses of neural networks for fluid flow estimation.
In the present paper, three perspectives which are remaining challenges for applications of machine learning to fluid dynamics are considered: 1. interpretability of machine-learned results, 2. bulking out of training data, and 3. generalizability of neural networks.
For the interpretability, we first demonstrate two methods to observe the internal procedure of neural networks, i.e., visualization of hidden layers and application of gradient-weighted class activation mapping (Grad-CAM), applied to canonical fluid flow estimation problems --- $(1)$ drag coefficient estimation of a cylinder wake and $(2)$ velocity estimation from particle images.
It is exemplified that both approaches can successfully tell us evidences of the great capability of machine learning-based estimations.
We then utilize some techniques to bulk out training data for super-resolution analysis and temporal prediction for cylinder wake and NOAA sea surface temperature data to demonstrate that sufficient training of neural networks with limited amount of training data can be achieved for fluid flow problems.
The generalizability of machine learning model is also discussed by accounting for the perspectives of inter/extrapolation of training data, considering super-resolution of wakes behind two parallel cylinders.
We find that various flow patterns generated by complex interaction between two cylinders can be reconstructed well, even for the test configurations regarding the distance factor.
The present paper can be a significant step toward practical uses of neural networks for both laminar and turbulent flow problems.
\end{abstract}

\section{Introduction}
A modern big wave of machine learning has propagated to fluid dynamics community.
In particular, neural networks, which have a great potential as an universal approximator \cite{Kreinovich1991,Hornik1991,Cybenko1989,BFK2018}, have acquired strong attentions from fluid mechanicians for various extensions \cite{BNK2020}.
Fundamental studies for closure modeling in large-eddy simulation (LES) and Reynolds Averaged Navier--Stokes (RANS) simulation can be regarded as one of enthusiastic topics in neural networks and fluid dynamics \cite{DIX2019}.
Gamahara and Hattori \cite{gamahara2017searching} applied a multi-layer perceptron (MLP) with only one hidden layer to LES closure and compared its ability to conventional models, considering a turbulent channel flow. 
An extension of a similar idea was performed by Maulik and San \cite{maulik2017neural} with a deconvolution approach.
Following these pieces of seminal work, various studies have tackled to neural network-based LES modeling so as to establish the universal closure that can be applied to a wide range of flows \cite{maulik2019sub,maulik2019subgrid,yang2019predictive,pawar2020priori}.
For RANS modeling, a notable work here is that of Ling et al. \cite{ling2016reynolds}.
They proposed a tensor-basis neural network (TBNN), which can guarantee a Galilean invariance, and tested the model for flows in a duct and over a wavy-wall.
The TBNN have been extended to various flow configurations \cite{milani2020turbulent} and more practical issues, e.g., uncertainty quantification \cite{geneva2019quantifying}.
In addition to the aforementioned supervised methods, Novati et al. \cite{novati2020automating} have recently proposed a reinforcement learning-based closure by considering a homogeneous isotropic turbulent flow, which enables us to expect new methods of machine learning-based turbulence modeling.

One of the outstanding characteristics in the neural network operation is the use of nonlinear activation functions.
It is widely known that neural networks can establish efficient reduced order models thanks to the nonlinearity caused herein \cite{THBSDBDY2020}.  
Wang et al. \cite{wang2018model} proposed a framework to predict the temporal evolution of proper orthogonal decomposition (POD) coefficients using the long short-term memory (LSTM) by considering an ocean gyre and a flow past a cylinder.
As an extension to turbulence, Srinivasan et al. \cite{SGASV2019} used the LSTM to predict the temporal evolution of the coefficients of the nine equation model for a turbulent shear flow and reported its great potential.
Focusing on spatial order reduction, neural network-based low dimensionalization, i.e., autoencoder (AE), is also one of the promising candidates \cite{Milano2002,FHNMF2020}.
The great role of nonlinear activation functions in neural networks was well summarized in Murata et al. \cite{MFF2019}, who compared AE-based modes to POD modes considering a laminar cylinder wake and its transient.
More recently, the customized AE referred to as a hierarchical AE was proposed by Fukami et al. \cite{FNF2020} to handle turbulent flows efficiently.

While information of high-resolution flow fields have allowed us to understand complex flow physics, uses of neural networks which account for nonlinearity into its regression procedure can also be found for data reconstruction and estimation \cite{FFT2020}.
For instance, Fukami et al. \cite{FNKF2019} used a combination of convolutional neural network (CNN) and MLP to predict the temporal evolution of a cross-sectional field in a turbulent channel flow and applied the unified model as an inflow turbulence generator.
The CNN-based model was also presented by Salehipour and Peltier \cite{SP2019} to predict the small scale motions in the ocean turbulence referred to as {\it atoms}.
From the perspective of image processing, super-resolution analysis, in which high-resolution data are recovered from its low-resolution counter part, was applied to turbulent flows by Fukami et al. \cite{FFT2019a,FFT2019b}.
The extension of this idea to higher Reynolds number flows \cite{LTHL2020}, experimental data \cite{DHLK2019}, and three-dimensional turbulence \cite{FFT2021b} can also be found.
In addition, a CNN-based velocity estimator for particle image velocimetry (PIV) was proposed by Cai et al. \cite{CZXG2019}. 
They examined its ability considering various flows.
The applicability of a similar method to deteriorated experimental images was recently investigated by Morimoto et al. \cite{MFF2020}.
To sum up, various methods for neural network-based fluid data enrichment were proposed for both numerical and experimental studies.

Furthermore, neural networks have played a significant role in the flow control community \cite{BHT2020}.
The first attempt of supervised machine learning-based flow control was performed in 1997 by Lee et al. \cite{lee1997application}.
Their model was trained to learn the control input of the opposition control \cite{CMK1994} using only the wall-sensor measurement so as to reduce the friction drag.
Many of recent efforts have been devoted to reinforcement learning \cite{garnier2019review}.
Rabault et al. \cite{rabault2019artificial} applied the reinforcement learning to perform active flow control with two jets on a cylinder surface.
The extension of the technique to different Reynolds numbers was assessed by Tang et al. \cite{tang2020robust}, which achieves significant drag reduction of 5.7\%, 21.6\%, 32.7\%, and 38.7\%, at $Re_D = 100, 200, 300$ and $400$, respectively.
Although these efforts are still limited to laminar flow cases, the success here motivates us its extension to turbulent flows.  

Although a wide range of neural network applications to fluid dynamics problems can be seen as introduced above, we still have some challenges toward more practical steps.
In this paper, we focus on three perspectives as follows:
\begin{enumerate}
    \item Interpretability of machine-learned results.\\
    In the practical sense, we should address the interpretability of results collected from machine learning, e.g., ground for estimations and uncertainty quantification. Also, in fluid dynamics fields, some researchers have tackled this issue: Maulik et al. \cite{MFRFT2020} have recently demonstrated the capability of probabilistic neural network (PNN) with a problem setting of POD coefficient prediction over a time of shallow water equation, vortex shedding behind a cylinder or an airfoil, and NOAA sea surface temperature. One of beauties in their work is that the PNN can tell us a confidence interval in estimating target attributes.
    Otherwise, Jagodinski et al. \cite{JZV2020} used a three-dimensional CNN with {gradient-weighted class activation mapping (Grad-CAM) \cite{selvaraju2016grad,selvaraju2017grad}} to identify the important area for {the} prediction of ejection events in a turbulent channel flow. In addition, Kim and Lee \cite{KL2020} examined the relationship between the estimation of the wall-normal heat flux and vortical motion of turbulent channel flow by looking inside CNN.
    \item Amount of training data.\\
    To extract the underlying physics of fluid flow data, massive amount of training data has been utilized for neural networks \cite{Kutz2017}. For example, Fukami et al. \cite{FFT2021b} reported that approximately 15 days are required to train their neural network to perform three-dimensional spatio-temporal super-resolution analysis. To reduce the computational cost and storage, a proper method to bulk out the training data while keeping the ability of neural networks is eagerly desired for the fluid dynamics problems.
    \item Generalizability of neural networks for fluid flows\\
    A generalized model beyond various kinds of flows is also one key factor toward next steps of machine learning and fluid dynamics. Some studies have recently examined this point to consider inter/extrapolation boundary in terms of training data. Hasegawa et al. \cite{HFMF2020b} examined the Reynolds number dependence in performing CNN-LSTM-based reduced order modeling considering a laminar cylinder wake. They also investigated the generalizability of geometric variation using the similar form of reduced order surrogate \cite{HFMF2020a}. Otherwise, Erichson et al. \cite{erichson2019} used {an MLP} referred to as {\it shallow decoder} to reconstruct fluid flows from local sensor measurements and discussed for inter/extrapolation of training data by considering two dimensional forced turbulence.
\end{enumerate}

The aim of the present paper is to demonstrate and introduce the capability of some techniques to clarify the aforementioned challenges in neural networks and fluid dynamics.
The main contribution of this paper is to investigate the applicability of various generalization techniques to high-dimensional nonlinear dynamics.
We cover various canonical neural network-based applications, i.e., force coefficient estimation, experimental velocity data estimation, spatial super resolution, and temporal prediction, using a wide range of fluid flow data and sea surface temperature data.
Although many machine learning studies have been conducted in physical science, the choice of used techniques highly depends on users' experience and intuition.
Hence, providing detailed analyses on the generalization techniques should be highly beneficial in a wide range of science and engineering.
The paper is organized as follows: the fundamental information on the fluid flow datasets covered in this study is provided in section \ref{sec:flow}.
The present machine learning models are introduced in section \ref{sec:mlm}.
As for the result part, we first introduce the visualization method inside neural networks in section \ref{sec:inside} with canonical regression problems.
The generalization techniques for the amount of training data and unseen data are then discussed in section \ref{sec:gen}.
Finally, concluding remarks are given in section \ref{sec:conclusion}.

\section{Flow fields used for training}
\label{sec:flow}

\begin{figure}
	\vspace{0mm}
	\centering
		\includegraphics[width=0.95\textwidth]{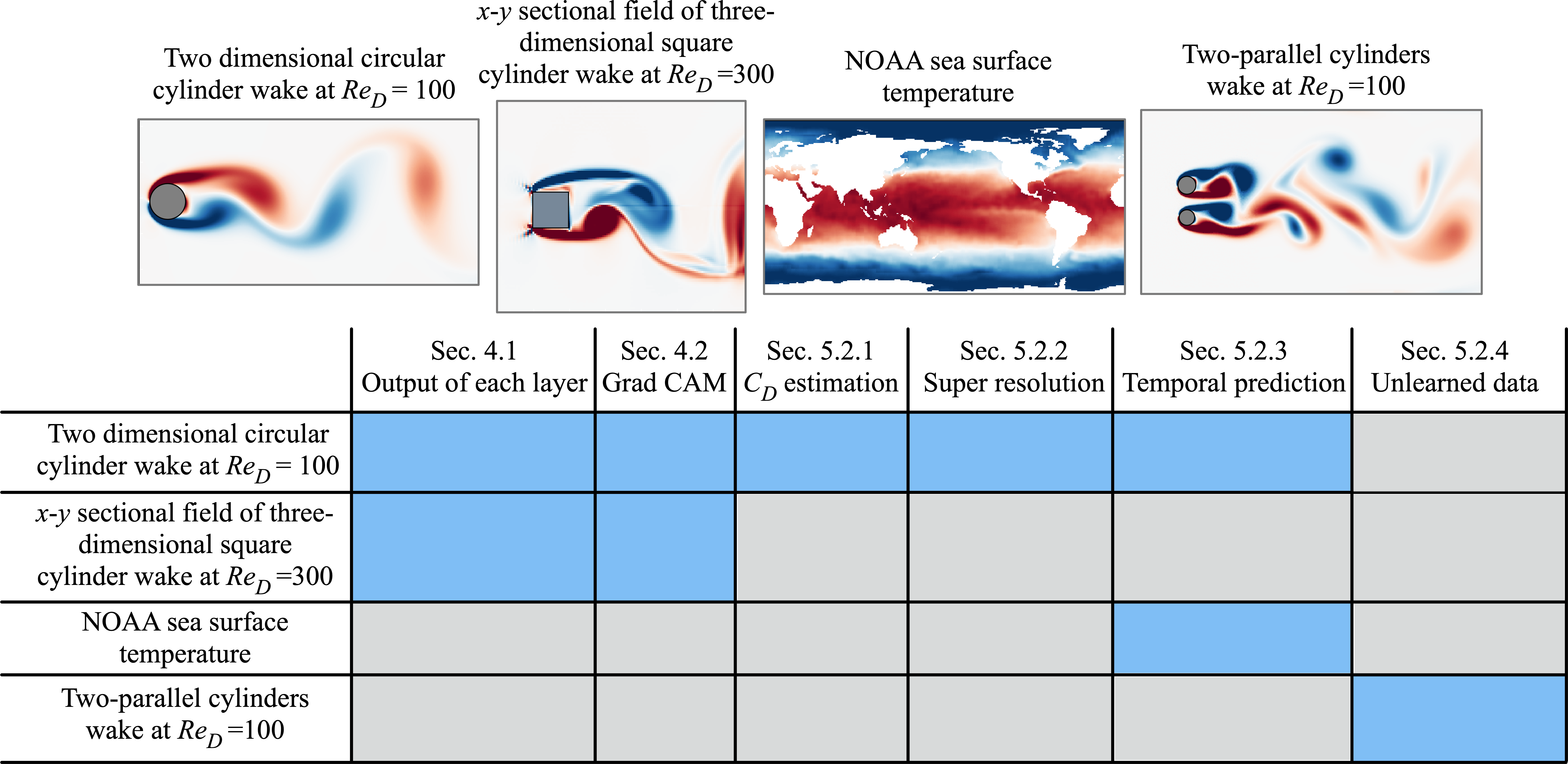}
		\caption{Flow fields used in the present study. In the table, blue boxes indicate the use of the flow field.}
		\label{fig_flowdata}
\end{figure}

We consider various flow fields to cover a wide range of complex fluid flow nature with several canonical problem settings, as summarized in figure \ref{fig_flowdata}.
In what follows, we introduce the setup for the training data used in this study. 

\subsection{Two-dimensional circular cylinder wake at $Re_D=100$}
\label{subsec:singleCylinder}

A temporally periodic wake behind a circular cylinder at ${Re}_D=100$ is mainly used for the demonstrations in this study.  
The datasets are generated with a two-dimensional direct numerical simulation (DNS). 
The governing equations are the incompressible continuity and Navier--Stokes equations, i.e.,
\begin{align}
    &\bm{\nabla} \cdot \bm{u}=0, \\
    &{\partial_t\bm{u}} + \bm{\nabla} \cdot (\bm{uu})  = - \bm{\nabla} p + \dfrac{1}{{Re}_D}\nabla ^2 \bm{u},
\end{align}
where $\bm{u}$ and $p$ denote the velocity vector and pressure, respectively.
All quantities are non-dimensionalized using the fluid density, the free-stream velocity, and the cylinder diameter.
The size of the computational domain is ($L_x, L_y$)=(25.6, 20.0), and the cylinder center is located at $(x, y)=(9,0)$.
The Cartesian grid system with the grid spacing of $\Delta x=\Delta y = 0.025$ is applied to the present simulation.
A no-slip boundary condition on the cylinder surface is imposed using an immersed boundary method \cite{kor2017}.
Although the number of grid points used for DNS is $(N_x, N_y)=(1024, 800)$, only the flow field around the cylinder is used as the training data whose dimension is $(N_x^*, N_y^*)=(384, 192)$ corresponding to a domain of $8.2 \leq x \leq 17.8$ and $-2.4 \leq y \leq 2.4$.
As for the data attributes, the vorticity field $\omega$ is considered. 
The time interval of flow field data is $\Delta t=0.25$, which corresponds to approximately 23 snapshots per period, with the Strouhal number of 0.172.

\subsection{A cross-sectional field of three-dimensional square cylinder wake at $Re_D=300$ and its particle images}

A flow around a square cylinder at the Reynolds number ${Re}_D=300$ is then used for our presentation with the machine learning-based PIV velocity estimator \cite{MFF2020} in section \ref{sec:inside}.
The training dataset is prepared by a DNS, which has been verified against Franke et al. \cite{FRS1990} and Robichaux et al. \cite{RBV1999}, with numerically solving the incompressible Navier--Stokes equations with a penalization term \cite{volumePenal1994},
\begin{align}
        &{\bm{\nabla}} \cdot {\bm u}=0, \\
        &{\partial_t {\bm u}} + {\bm{\nabla}} \cdot \left({\bm{u}}{\bm{u}}\right)=-{\bm{\nabla}} p + \frac{1}{{ Re}_D}{\bm{\nabla}}^2\bm{u}+\lambda \chi\left({\bm u}_b-{\bm u}\right),\\ \nonumber
\end{align}
where the penalization term, which represents an object, is expressed with a penalty parameter $\lambda$, a mask value $\chi$, and a velocity vector of a flow inside the object ${\bm u}_b$, which is zero for the fixed object.
The mask value is $\chi=0$ in the flow domain and $\chi=1$ inside the object.
The size of the computational domain here is {$\left(L_x, L_y, L_z\right)=\left(20D,20D,4D\right)$.}
{The computational time step is set to $\Delta t=5.0\times10^{-2}$.}
For the training data of PIV example, we focus on the volume around the square cylinder, i.e., {$\left(7D\times6D\times0.5D\right)$.}
The number of grid points of the extracted region is $(N_x^*, N_y^*, N_z^*)=(140,120,20)$.
To consider the three-dimensionality of the present flow at ${Re}_D=300$ \cite{BA2018}, twenty $x-y$ cross-sections at different {spanwise} locations are used for the training data.
The details of preparation for particle images can be found in Morimoto et al. \cite{MFF2020}.

\subsection{NOAA sea surface temperature}

The NOAA sea surface temperature dataset \cite{noaa}, obtained from satellite and ship observations, is used to examine the behavior of machine learning models in practical situations which have no modelled governing equations, e.g., geophysical observation.
We here use the weekly observation data, which comprise of a spatial resolution of $360\times180$ based on a $1^\circ$ grid.
In the present study, we insert zero for the continental portions, which are colored by white in figure \ref{fig_flowdata} for clarity of illustration.

\subsection{Two-parallel cylinders wake at $Re_D=100$}
\label{subsection:2Psetup}

\begin{figure}
    \centering
    \includegraphics[scale=0.5]{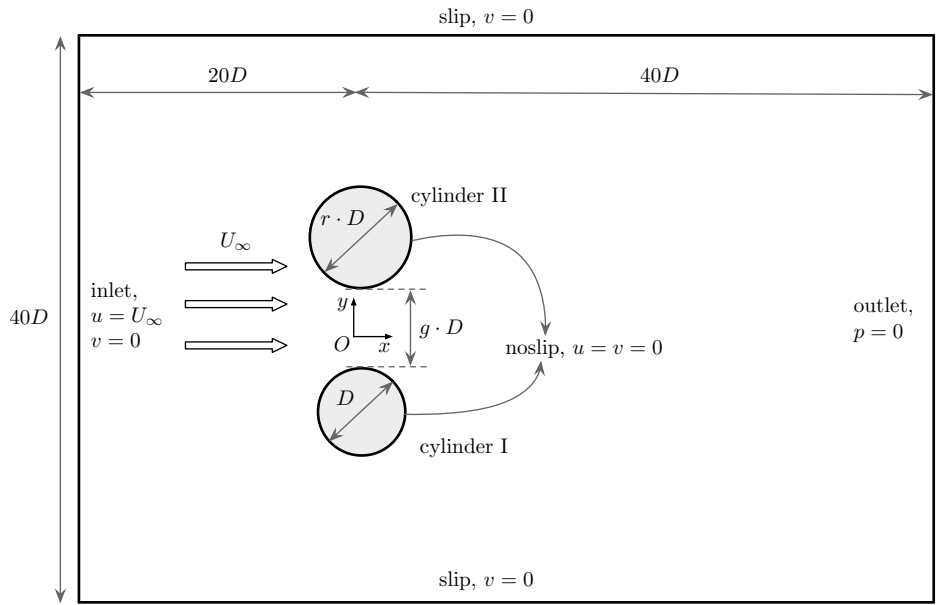}
    \caption{Computational setup for flow over two side-by-side cylinders.}
    \label{fig:2PScheme}
\end{figure}

A more complicated flow comprising the wake interactions between two side-by-side uneven circular cylinders is also considered to 
discuss the boundary of inter/extrapolation for training data. 
A schematic view of the problem setup is shown in figure \ref{fig:2PScheme}. 
The two circular cylinders with a size ratio of $r$ are separated with a gap of $gD$, where $g$ is the gap ratio. 
The Reynolds number is fixed at $Re_D=U_{\infty}D/\nu=100$. 
The two cylinders are placed $20D$ downstream of the inlet where a uniform flow with velocity $U_{\infty}$ is prescribed, and $40D$ upstream of the outlet with zero pressure. 
The side boundaries are specified as slip and are $40D$ apart. 
The flows over the two cylinders are solved by the open-source CFD toolbox OpenFOAM \cite{weller1998tensorial}, using second-order discretization schemes in both time and space.

\begin{figure}
	\centering
	\includegraphics[width=0.9\textwidth]{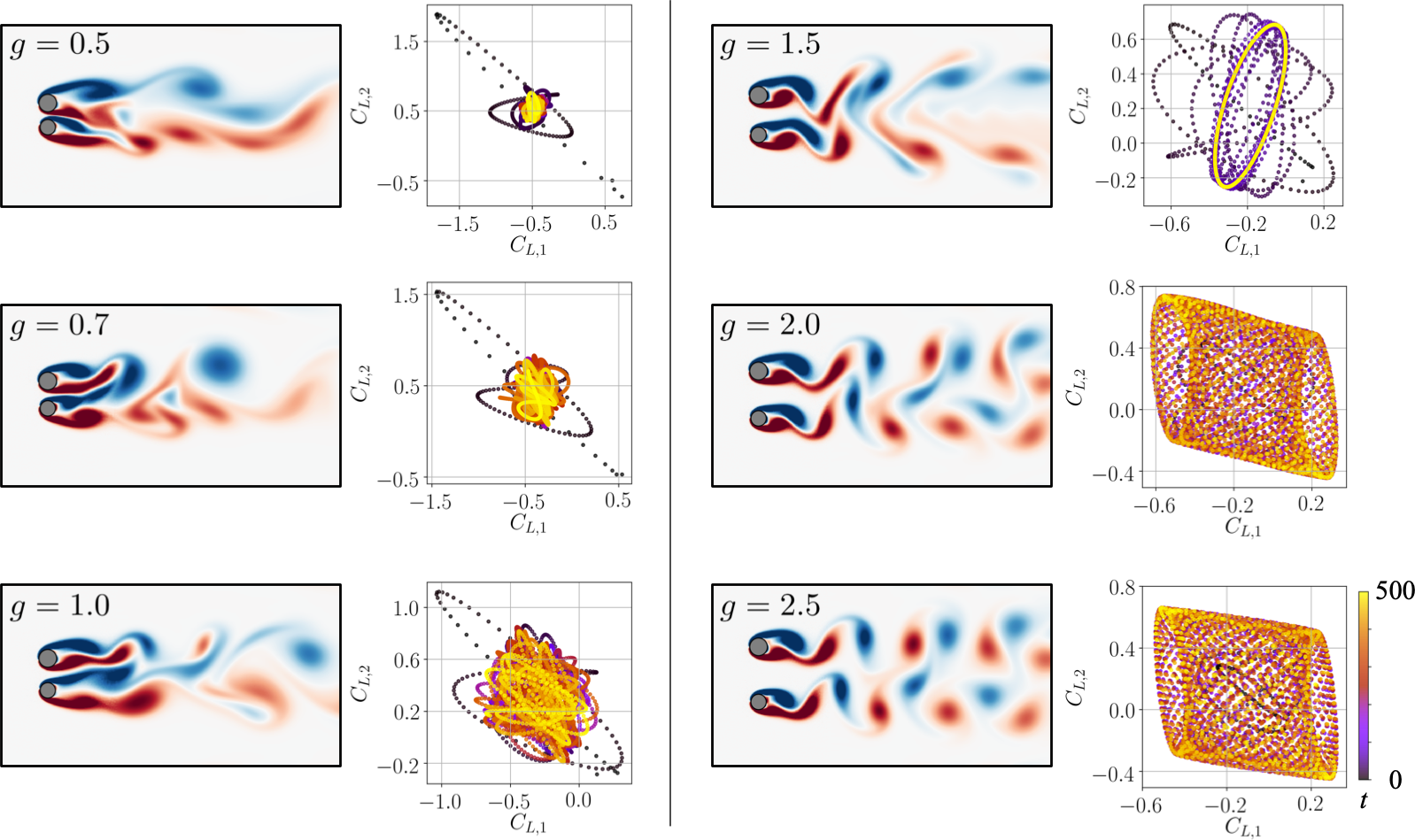}
	\caption{Vorticity fields with Lissajous plot for $C_{L,1}$ and $C_{L,2}$ of two-parallel cylinders wake.}
	\label{fig_2cy_CL}
\end{figure}

As the size ratio $r$ and gap ratio $g$ are varied, the flow over the two cylinders exhibits various wake patterns, which will be discussed in more detail in our coming paper. 
For the present study, we fix the size ratio $r$ to 1.15 and vary the gap ratio $g$ from 0.5 to 2.5. 
The wake patterns and the corresponding Lissajous plots ($C_{L,1}$--$C_{L,2}$) are shown in figure \ref{fig_2cy_CL}.
At low gap ratios ($g=0.5, 0.7$ and 1.0), the wakes are characterized by irregular interactions of the two vortex streets. 
The phase spaces spanned by the two lift coefficients feature their chaotic trajectories. 
As the gap ratio is increased to $g=1.5$, the two vortex streets in the near wake merge into one in far wake. 
This is also accompanied by the frequency lock-in among the two nonlinear oscillators. 
Further increasing the gap ratio to $g=2.0$ and 2.5, the vortex shedding of the two cylinders takes place independently with their respective natural frequencies, and the wake is featured by complex vortex interactions. 

\section{Machine learning models}
\label{sec:mlm}

Machine learning models used in the present study are constructed by a multi-layer perceptron (MLP) and/or convolutional neural network (CNN).
We consider several combinations of them depending on the problem setting, i.e., the size of dimension of the handled data.
Here, let us briefly introduce the fundamental theories of MLP (section \ref{sec:mlp}) and CNN (section \ref{sec:cnn}), then explain the present models, which are comprised of them in section \ref{sec:comb}.

\subsection{Multi-layer perceptron}
\label{sec:mlp}

A multi-layer perceptron (MLP) \cite{RHW1986} was inspired by the structure of biological neural circuits.
The MLP has successfully been applied to not only computer science field \cite{Domingos2012}, but also fluid dynamics community for turbulence modeling, reduced order modeling, and data estimation \cite{LW2019,MSRV2019,YH2019}.

\begin{figure}
	\centering
	\includegraphics[width=0.7\textwidth]{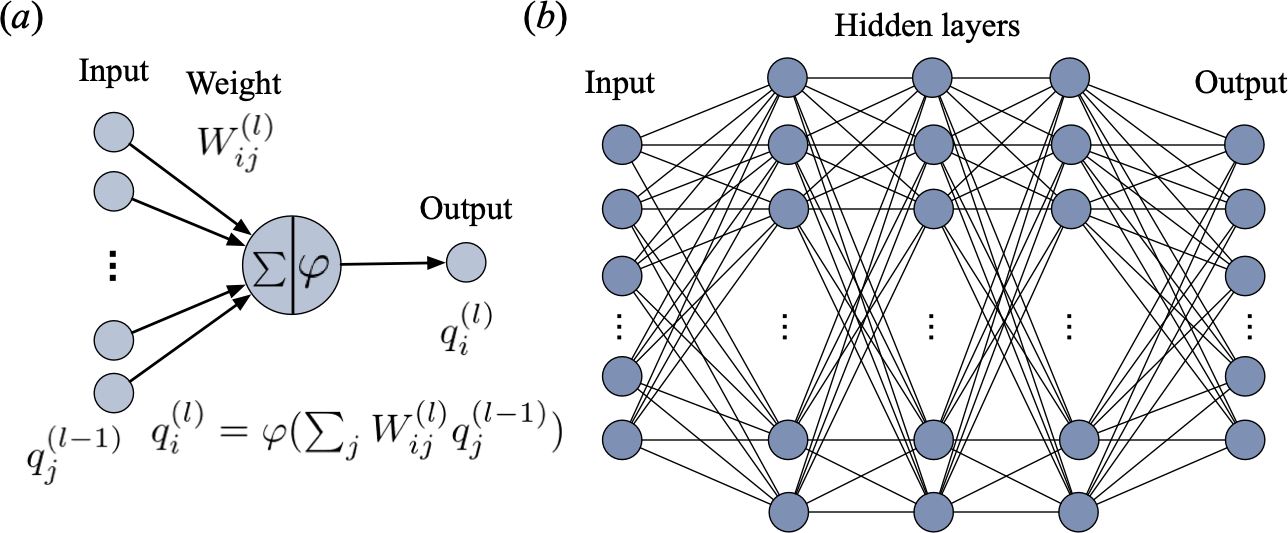}
	\caption{
	$(a)$ A perceptron. $(b)$ Multi-layer perceptron is an aggregate of perceptrons.}
	\label{fig_mlp}
\end{figure}

The MLP illustrated in figure \ref{fig_mlp}$(b)$ is constructed by a lot of perceptrons, which is a minimum unit as shown in figure \ref{fig_mlp}$(a)$. 
In the perceptron, the {output} data at the $(l-1){\rm th}$ layer are fed into the next layer $(l)$ while taking a weight $\bm{W}$. 
Notable characteristics of MLP is that the linear superposition is then passed through a nonlinear activation function $\varphi$ such that 
\begin{eqnarray}
    q_i^{(l)}=\varphi\left(\sum_j W^{(l)}_{ij}q_j^{(l-1)}\right).
    \label{eq:1}
\end{eqnarray}
Weights on all connections $W_{ij}$ are optimized to minimize a cost function ${E}$ with a back propagation \cite{KB2014}, i.e., ${\bm w}={\rm argmin}_{\bm w}[{E}({\bm y},{\mathcal F}({\bm x};{\bm w}))]$.
In the MLP formulation, we use ReLU activation function \cite{NH2010}, which works well for weight update issue of deep neural networks.

\subsection{Convolutional neural network}
\label{sec:cnn}

Convolutional neural networks (CNN) \cite{LBBH1998} have mainly been utilized for image recognition tasks since the filter operation inside the CNN enables us to handle high-dimensional data without encountering the curse of dimensionality.  
The capability of CNN has also encouraged uses of CNN in the field of fluid dynamics \cite{FNKF2019,SP2019,matsuo2021supervised,2021wallmodel,nakamura2021MLLSE}.

\begin{figure}[b]
	\centering
	\includegraphics[width=0.80\textwidth]{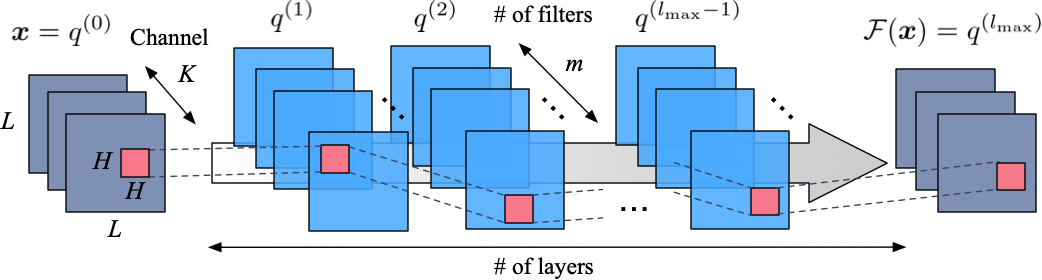}
	\caption{Two-dimensional convolutional neural network.}
	\label{fig-cnn}
\end{figure}

In this study, we use a two-dimensional CNN illustrated in figure \ref{fig-cnn}.  
An input data ${\bm x}=q^{(0)}$, which has $L\times L$ pixels, is fed into the layer $(l)$ -- and then repeating manner from layer $q^{(l-1)}$ to layer $q^{(l)}$, where $l~(0\leq l \leq l_{\rm max})$, is applied.
The procedure of CNN for $q^{(l)}$ can be mathematically given as 
\begin{eqnarray}
    q^{(l)}_{ijm} = {\varphi}\biggl(b_m^{(l)}+\sum^{K-1}_{k=0}\sum^{H-1}_{p=0}\sum^{H-1}_{s=0}h_{p{s}km}^{(l)} q_{i+p-C,j+{s-C},k}^{(l-1)}\biggr),
\end{eqnarray}
where {$C={\rm floor}(H/2)$,} $b_m^{(l)}$ is the bias, $q^{(l_\text{max})} = {\mathcal F}({\bm x})$, and $K$ is the number of {input data channels}. 
The pink square of $H\times H$ in figure \ref{fig-cnn} represents the filter $h$. 
Similar to the weight update in MLP formulated as equation (\ref{eq:1}), weights in CNNs, i.e., the filtering coefficients, are also obtained through an optimization manner.
As the filter of size $H\times H$ is shared for whole image of $L\times L$, the filter operation in CNN is generally called weight sharing, which allows us to handle big data with significantly lower computational costs compared to the fully-connected MLP.  
A max pooling layer is utilized for dimensional reduction.
Also, an up-sampling layer, which copies the values in the lower dimensional space into a higher dimension, is applied for dimensional extension.

\subsection{Unified models in this study with problem setting}
\label{sec:comb}
\begin{figure}
	\centering
	\includegraphics[width=0.8\textwidth]{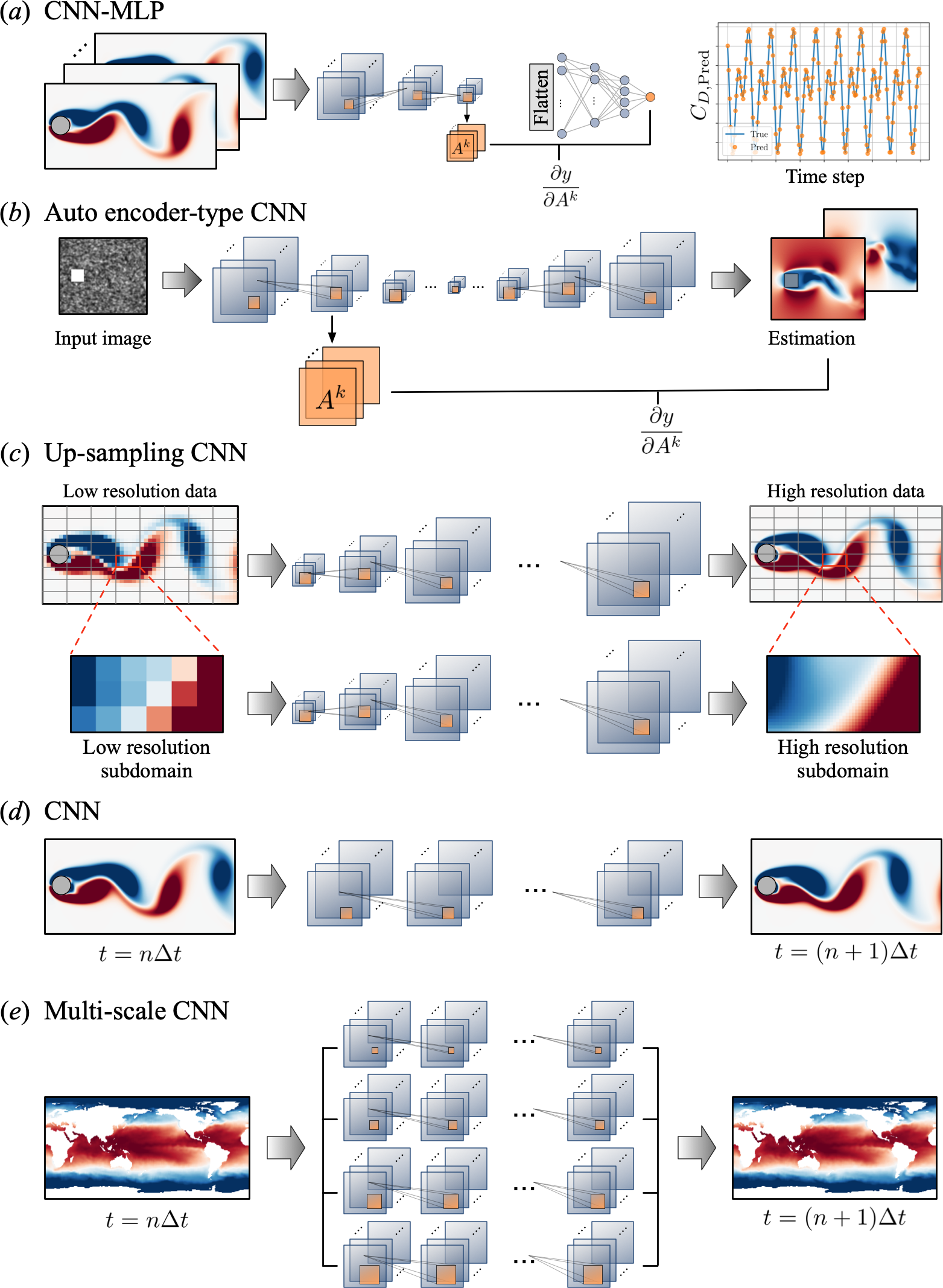}
	\caption{Problem settings with machine learning models in this study. $(a)$ CNN-MLP model used for scalar output example. $(b)$ Autoencoder-type CNN used for PIV velocity estimation. $(c)$ Up-sampling CNN for super-resolution analysis. $(d)$ Regular CNN for the temporal prediction of cylinder wake. $(e)$ Multi-scale CNN for the temporal prediction of NOAA sea surface temperature.}
	\label{fig_schm}
\end{figure}

{We use some different structures of neural network suited for each problem setting in this study.
Each model is comprised of convolutional neural network (CNN) and/or multi layer perceptron (MLP).
Problem settings and illustration of models covered in this study are summarized in figure \ref{fig_schm}.

For the visualization inside machine-learned models (sec. \ref{sec:inside}) and the demonstration of data bulking techniques (sec. \ref{sec:cdbulk}), the drag coefficient $C_D$ estimation for two-dimensional cylinder wake at $Re_D=100$ is performed \cite{FFT2020}.
Since the input--output relationship here is a two-dimensional input with a scalar output as can be seen in figure \ref{fig_schm}$(a)$, we first capitalize on the CNN with the pooling operations and then the MLP is inserted for low-dimensional vectors.

We also consider an experimental data estimation for PIV images, which is described in our previous work \cite{MFF2020}, to demonstrate the visualization techniques inside machine learned models.
As illustrated in figure \ref{fig_schm}$(b)$, an {autoencoder (AE)-type CNN} is trained to estimate a velocity field from the corresponding particle image.
Note in passing that we adopt the AE-like structure, which is robust for noise and spatial sensitivity, to meet the requirement in handling experimental images properly \cite{MFF2020}.

To examine the possibility of a data augmentation technique, we also consider super-resolution analysis \cite{FFT2019a}.
We here examine two methods of training, global data and local data, as shown in figure \ref{fig_schm}$(c)$.
We utilize the same network structure for both training processes, which is an up-sampling-based CNN.
In the model, the input low-resolution data is extended up to the dimension in which it matches with the data size of high-resolution data, and then the data are convolved to output section.

We also consider the temporal prediction of cylinder wake and NOAA sea surface temperature in this study using CNN without any pooling or up-sampling layers, since dimensions of both input and output are same as each other.
The model utilized for temporal prediction of cylinder wake is illustrated in figure \ref{fig_schm}$(d)$.
We use only the convolutional layers with filter size of $(5\times5)$.
In contrast, for the NOAA SST dataset, we utilize a multi-scale CNN \cite{DuMSCNN2018} since we can suspect that the flow field contains a wide nature of scales.
In our preliminary test, we have checked that the regular CNN does not work for the NOAA SST data, while the multi-scale CNN performs well.
As presented in figure \ref{fig_schm}$(e)$, the multi-scale CNN used in this study includes four different filter sizes, i.e., $(3\times3),\ (5\times5),\ (7\times7)$ and $(9\times9)$.
}

\section{Observing internal procedure of machine learning models}
\label{sec:inside}

Considering the practical uses of neural network for various purposes, the interpretability is one of the significant requirements.
Since the internal states of neural networks can be visualized with some techniques, we can expect that we may be able to find some physical insights or evidence of their estimations by observing the internal procedure.
Here, let us introduce two methods in observing the inside neural networks, i.e., visualization of hidden layers (section \ref{sec:layer-vis}) and gradient-weighted class activation mapping \cite{JZV2020} (section \ref{sec:grad-cam}).

\subsection{Output of each hidden layer in convolutional neural networks}
\label{sec:layer-vis}
\begin{figure}
	\centering
	\includegraphics[width=0.8\textwidth]{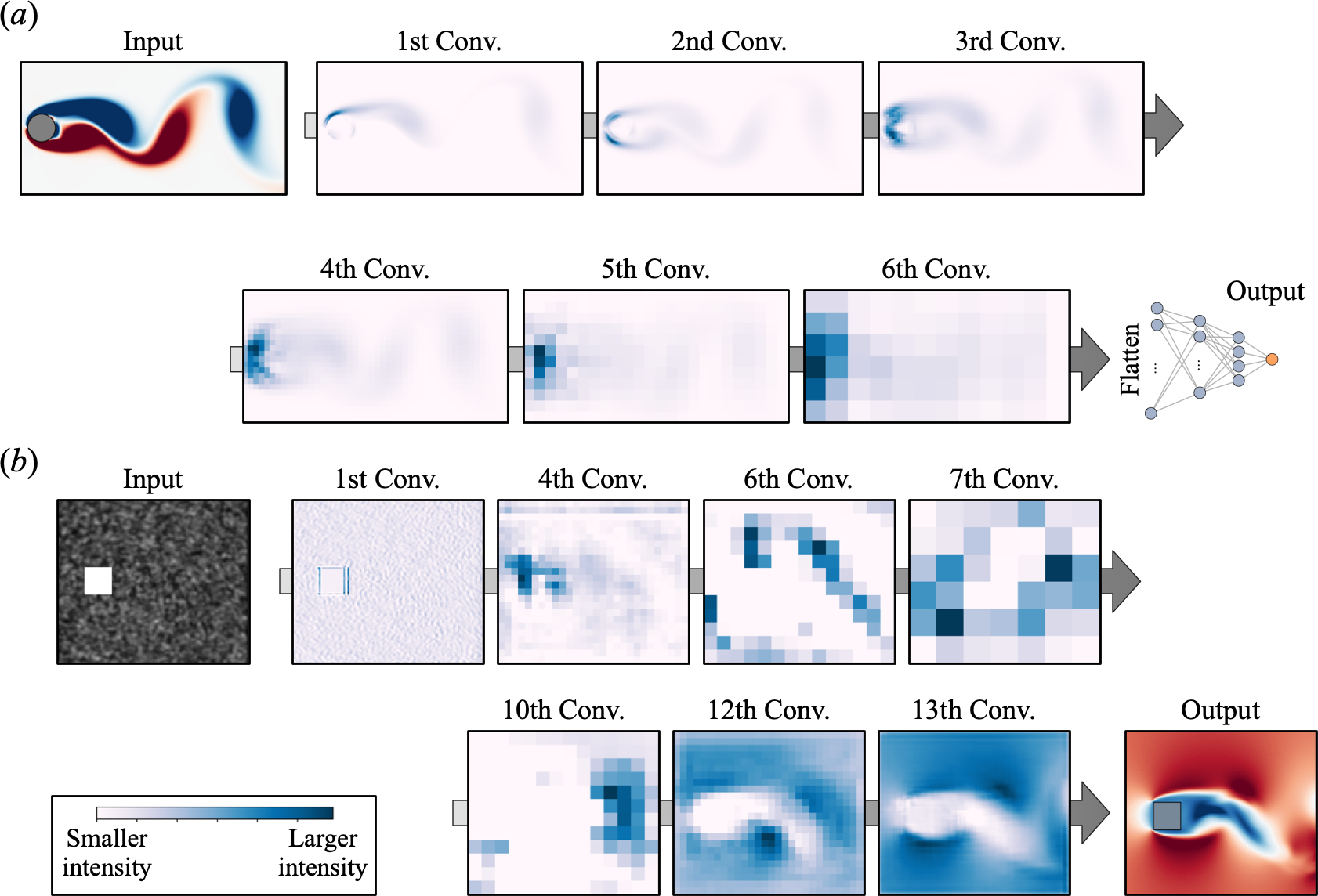}
	\caption{Output of each convolutional layer of $(a)$ machine learning-based $C_D$ estimation and $(b)$ PIV velocity estimation.}
	\label{fig_layer_vis}
\end{figure}

As the first technique for visualizing inside the model, we focus on hidden layers of neural networks.
As an example, let us present in figure \ref{fig_layer_vis} the output of each convolutional layer of {drag coefficient estimation \cite{FFT2020} and} experimental data estimation \cite{MFF2020}.
These are canonical problem settings in fluid dynamics.
Since a force coefficient estimation plays a crucial role in fluid engineering, there is a demand in obtaining the force coefficients in less computational cost, i.e., without numerically solving the flow.
For instance, Zhang et al.~\cite{ZSM2018} utilized a CNN to estimate a force coefficient of an airfoil from flow characteristics and geometry information.
Similar studies can also be found in Refs.~\cite{FFT2020,MJ2018}.
In contrast, neural network-based experimental velocity data estimation addresses the various experimental constraints, e.g., obtaining denser flow motion~\cite{CZXG2019}, and inpainting the data missing region of an experimental image~\cite{MFF2020}.

As presented in section \ref{sec:comb} and figure \ref{fig_schm}$(a)$, the CNN-MLP model is applied to the estimation of drag coefficient $C_D$.
The convolutional part of the model, which comprises of 6 convolutional layers and 5 pooling layers, is first utilized and then the MLP part with 4 hidden layers is combined for the scalar output.
As the representative output at each layer, we only extract the hidden output which reports largest average value over all channels per a layer based on the assumption that the outputs with larger weights have a significant contribution for the estimation, as presented in figure \ref{fig_layer_vis}.
As explicitly shown in figure \ref{fig_layer_vis}$(a)$, the region around the cylinder has {the larger intensity} at every layer, which is reasonable since the drag coefficient is determined by the variables on the cylinder surface.

For the task of experimental data estimation, the CNN model attempts to output a velocity field from artificial particle image as is detailed in Morimoto et al. \cite{MFF2020}.
As stated in section \ref{sec:comb}, we use an autoencoder (AE)-like CNN which comprises of 14 convolutional layers and 6 pooling/up-sampling layers.
In this case, it is able to estimate the role of each layer by extracting the output value of each hidden layer.
For the upper-stream layers, e.g., the 1st and 4th convolutional layers, the region around the square cylinder has larger values.
On the other hand, it is obvious that the downstream layers, e.g., the 12th and 13th layers, have relatively large intensities on the region with velocity fluctuation.
It implies that the upper-stream layers are responsible for recognizing the alignment of bluff body while the downstream layers attempts to output the velocity fluctuations.

\subsection{Gradient-weighted class activation mapping (Grad-CAM)}
\label{sec:grad-cam}

As we demonstrated in the previous section, we can estimate a role of each layer by observing the outputs at each hidden layer.
However, the weakness of the method is that it is unclear whether the large intensity output directly represents the contribution for the estimation and we need to speculate the meaning of each output field.
As for more interpretable tool to observe the internal procedure, we here apply a gradient-weighted class activation mapping (Grad-CAM) \cite{selvaraju2016grad,selvaraju2017grad} to our problem settings.
A Grad-CAM has been widely used on the field of image recognition, thanks to its capability in telling us the region with higher interest for the estimation of the trained model.
Other than the image classification, Jagodinski et al. \cite{JZV2020} has recently demonstrated the ability of the Grad-CAM for the probability estimation of ejection events in a turbulent channel flow.
In our study, the ability of Grad-CAM is demonstrated with canonical regression problems, i.e., $C_D$ estimation of a two-dimensional cylinder wake \cite{FFT2020} and experimental data estimation \cite{MFF2020} as shown in figures \ref{fig_schm}$(a)$ and $(b)$.
For the problem setting of experimental data estimation, we particularly choose the machine learned model trained by artificial particle images with data lacked region.
The model is originally trained to estimate the velocity field from lacked artificial particle images of flow around both single and double square cylinders.
For more details on this procedure, we refer readers to Morimoto et al. \cite{MFF2020}.

Basis of the Grad-CAM is a calculation of the gradient of output and designated convolutional layer.
For image classification problems, the model is generally consisted with convolutional layers and multi-layer perceptrons \cite{VGG16}.
To obtain an intensity map of the interest, the gradient between the output value and last convolutional layer which corresponds to the layer just before the multi-layer perceptron, needs to be calculated.
The model used in the $C_D$ estimation of a cylinder wake has a similar structure to those image classification networks since the output is a scalar value as presented in figure \ref{fig_schm}$(a)$.
Hence, we use the gradient $\partial y/\partial A_{ij}^k$ just as the classification problems, where $y$ denotes the estimated $C_D$ value and $A_{ij}^k$ explains the output of channel $k$ at the last convolutional layer, respectively.
The weight $\alpha^k$ for each channel can be then obtained taking the average of the gradients for each value as, 
\begin{equation}
    \alpha^k = \frac{1}{Z}\sum_i\sum_j\frac{\partial y}{\partial A^k_{ij}},
\end{equation}
where $(i,j)$ is the index of field $A_{ij}$ and $Z$ is a number of dimension of output field.
We then get the the Grad-CAM map $L$ as a superposition of weighted channels,
\begin{equation}
    L_{ij}={\rm ReLU}\left(\sum_k\alpha^k A^k_{ij}\right).
\end{equation}
Note that the ReLU function is applied here in order to consider only the positive influence.
As for the PIV data estimation, whose CNN model has a two-dimensional input-output relationship as shown in figure \ref{fig_schm}$(b)$, we examine the gradient calculation on the first and the last convolutional layers to observe the difference of the role of each layer.

\begin{figure}
	\centering
	\includegraphics[width=0.8\textwidth]{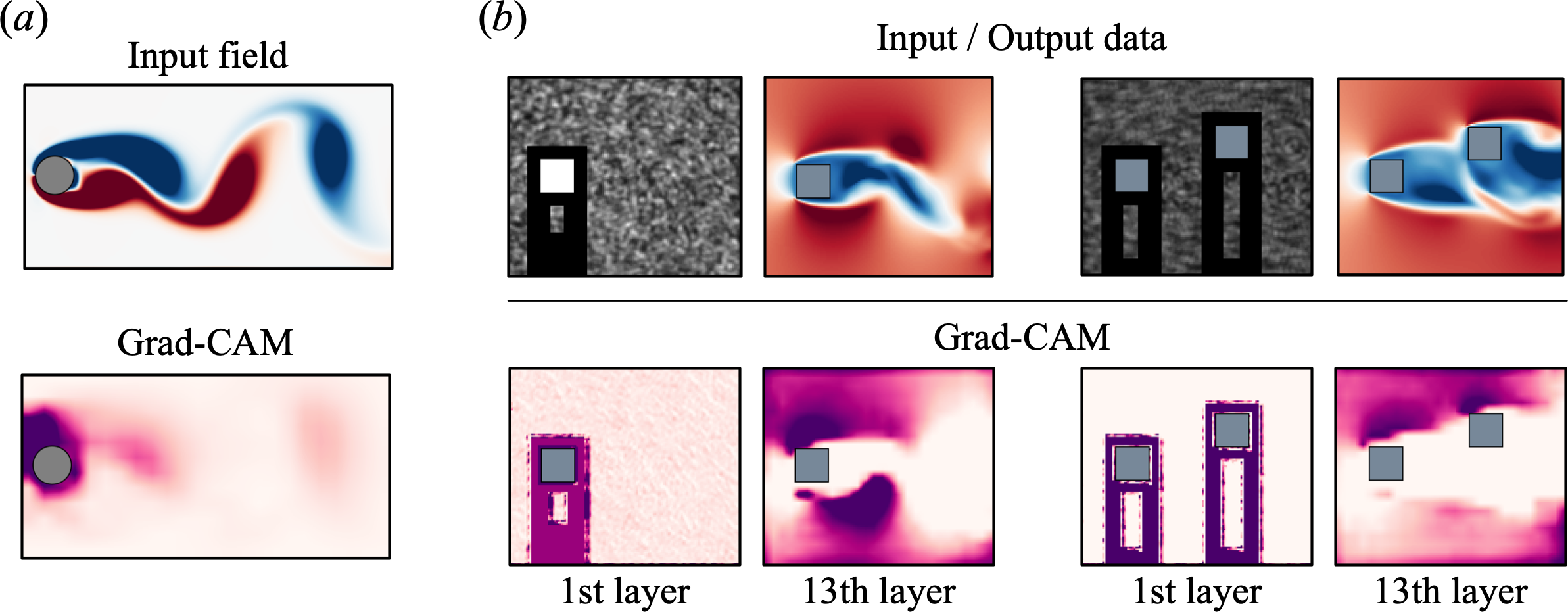}
	\caption{Grad-CAM-based visualization. $(a)$ $C_D$ estimation of two-dimensional cylinder wake. $(b)$ PIV velocity estimation for both single and staggered square cylinders.}
	\label{fig_grad-cam}
\end{figure}

Let us present in figure \ref{fig_grad-cam} the Grad-CAM maps $L$ of both problem settings.
As shown in figure \ref{fig_grad-cam}$(a)$, the Grad-CAM map indicates that the region around body is highly responsible for the $C_D$ estimation --- this observation is similar to the result of hidden layer visualization.
For the experimental data estimation shown in figure \ref{fig_grad-cam}$(b)$, the observable trend is also akin to the layer visualization in figure \ref{fig_layer_vis}, that the upper- stream layer has higher interest on the body alignment and the downstream layer is responsible for the velocity fluctuation --- but notable here is its clarity compared to the layer visualization.
The Grad-CAM map at the first layer shows that the network is obviously recognizing the alignment of square cylinders to classify the flow type between single and double square cylinder flow.
As demonstrated here, the Grad-CAM can be a simple but the powerful tool for observing the grounds of the estimation and it can be applied to not only classification problems but also regression problems, which are common in various demands of fluid dynamics community.

\section{Generalization technique for machine learning models}
\label{sec:gen}

In this section, we apply several well-known methods of training data bulking to canonical fluid flow problems to achieve a low-error level with a small amount of training data.
In what follows, the covered methods are briefly introduced with expected benefits for each problem setting.

\subsection{Method}
\label{sec:method}

\subsubsection{Flip in horizontal axis}
\label{sec:fliph}

\begin{figure}
	\centering
	\includegraphics[width=0.5\textwidth]{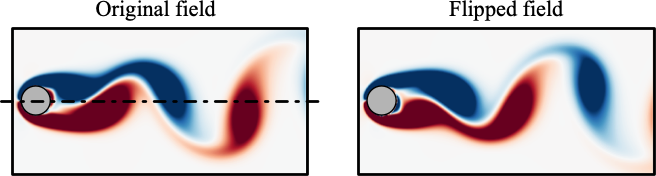}
	\caption{Original and flipped data of two-dimensional cylinder wake. The dotted line located on center region in the original field is a flipping axis.}
	\label{fig_flip}
\end{figure}

One of the simplest techniques to increase the amount of training data is to flip the field around proper axis.
Hasegawa et al. \cite{HFMF2020a} used the flipping technique for their training data, which is a laminar periodic shedding behind a bluff body.
Analogous to this study, we also apply the technique to the cylinder wake as shown in figure \ref{fig_flip}.
Since the cylinder wake data is statistically {symmetrical} with the horizontal axis, we can simply get the snapshot of half-cycle ahead (or ago), meaning that we could double the number of {training snapshots such that,
\begin{equation}
    n_{\rm flip} = 2n_{\rm DNS},
\end{equation}
where $n_{\rm DNS}$ is the amount of the reference DNS data and $n_{\rm flip}$ is the number of overall training snapshots obtained through data flipping, respectively.}
Note that users have to care {\it ergodicity} depending on the target dataset and flipping axis \cite{guastoni2020convolutional}, i.e., the statistical features of flipped data and original DNS data are common with each other in this particular example.

\subsubsection{Noise addition}
\label{sec:noiseadd}

Another feasible technique is to add noise to the training data.
{Neural networks can be generalized, i.e., to avoid overfitting, by utilizing non-physical noisy measurement, since the test data can be generally regarded as `noisy' data against the training data \cite{shorten2019survey}.}
We can increase the training data infinitely such that,
\begin{equation}
    n_{\rm noise} = \gamma n_{\rm DNS},\ \gamma=2,3,4,\cdots,
    \label{eq_noise}
\end{equation}
where $\gamma$ is an increasing rate of data amount and $n_{\rm noise}$ is the number of training snapshots increased by random noise addition.
Although several kinds of artificial noises can be considered \cite{HLQW2020}, e.g., Gaussian, speckle, and salt \& pepper, for training data, we here use the uniformly distributed random noise among $-0.1$ to $0.1$ for bulking out the training data as an example.

\subsubsection{Transfer learning with spatial local data}
\label{sec_tl}

\begin{figure}
	\centering
	\includegraphics[width=0.5\textwidth]{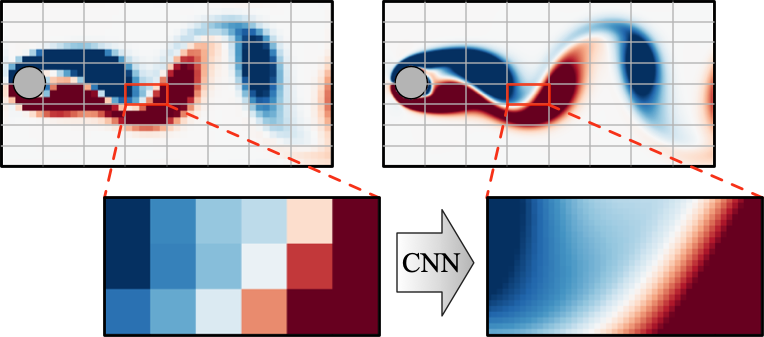}
	\caption{Zoom-in concept of two-dimensional cylinder wake for spatial super-resolution reconstruction. The original domain is meshed into 64 sub-domains.}
	\label{fig_sr_schm}
\end{figure}

Another well known technique to bulk out the training data in image recognition is to zoom-in and/or -out the training image \cite{mikolajczyk2018data}.
{We here borrow the zoom-in/out concept for the fluid flow estimation.
In this study, for the demonstration of super-resolution analysis, we combine the concept of zoom-in augmentation to transfer learning (or fine tuning), which has also been known as a good candidate to ease the training process by setting proper initial weights \cite{guastoni2020convolutional}.}
To prepare the zoomed-in image, we simply divide the original training data into several sub-domains as shown in figure \ref{fig_sr_schm}.

Supervised transfer learning process utilized in this study can be implemented as follows.
We first train the neural network ${\cal F}_{\rm pre}$ with local sub-domains $\bm{q}_{\rm local}$,
\begin{equation}
    {\bm q}_{\rm Out,local}={\cal F_{\rm pre}}({\bm q}_{\rm In,local};{\bm w}_{\rm pre}),
\end{equation}
where the subscripts Out and In stand for output and input data realizations, respectively.
Since we will consider the super-resolution analysis for transfer learning, the local model ${\cal F}_{\rm pre}$ learns the relationship between local low-resolution data and local high-resolution counterpart, as shown in figure \ref{fig_sr_schm}.
The weights ${\bm w}_{\rm pre}$ obtained through a minimization manner, ${\bm w}_{\rm pre}={\rm argmin}_{{\bm w}_{\rm pre}}||{\bm q}_{\rm Out,local}-{\cal F_{\rm pre}}({\bm q}_{\rm In,local};\bm{w}_{\rm pre})||_2$, will be then set as initial weights of posteriori network ${\cal F}_{\rm post}$.
Hence, the training process can be mathematically written as
\begin{eqnarray}
    &&{\bm w}_{\rm post}={\rm argmin}_{{\bm w}_{\rm post}}||{\bm q}_{\rm Out,global}-{\cal F_{\rm post}}({\bm q}_{\rm In,global};\bm{w}_{\rm post})||_2,
\end{eqnarray}
{where} ${\bm w}_{\rm post,init}={\bm w}_{\rm pre}$.
Again, we can expect the improvement of reconstruction accuracy through the learning process for both local and global fields.

\subsection{Demonstration}

\subsubsection{$C_D$ estimation of a flow around cylinder}
\label{sec:cdbulk}

\begin{figure}
	\centering
	\includegraphics[width=\textwidth]{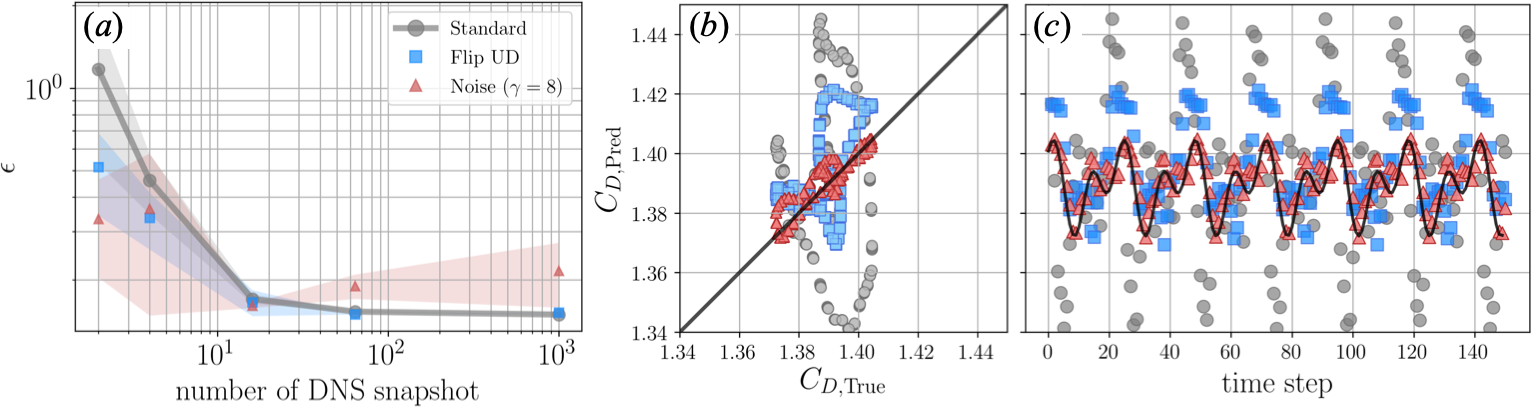}
	\caption{$C_D$ estimation of cylinder wake at $Re_D=100$. $(a)$ Dependence of $L_2$ error norm on number of training DNS snapshots for each data augmentation. $(b)$ True-predicted plot and $(c)$ time series of $C_D$ at ${n_{\rm snapshot}}=2$.}
	\label{fig_cd_result}
\end{figure}

Here, let us demonstrate the data bulking techniques by considering the drag coefficient estimation of two-dimensional cylinder wake.
We cover two methods for this $C_D$ estimation: 1. data flipping (sec. \ref{sec:fliph}) and 2. noise addition (sec. \ref{sec:noiseadd}).
Note that we skip the use of transfer learning since it was clearly observed in the previous section that the region around cylinder is highly responsible for the accuracy of estimation, which indicates that sub-domain meshing is likely not helpful for this problem setting.
The numbers of original snapshots used for training $n_{\rm DNS}$ are set to $n_{\rm DNS}=\{2,4,16,64,1000\}$ so as to investigate the dependence of estimation ability on number of training snapshots.
Hence, the numbers of bulked out data via data flipping are $n_{\rm flip}=2\times n_{\rm DNS}=\{4,8,32,128,2000\}$.
In addition, we set the increasing rate $\gamma$ in equation \ref{eq_noise} to $\gamma=8$ for the noise-based data augmentation such that $n_{\rm noise}=8\times n_{\rm DNS}=\{16,32,128,512,8000\}$.

The $L_2$ error norm of estimated $C_D$ for the covered bulking techniques is shown in figure \ref{fig_cd_result}$(a)$.
Note that the horizontal axis is arranged by the number of the original DNS data used for the training process $(n_{\rm DNS})$ to check the influence on each bulking technique.
Hence, the actual numbers of snapshots used for a training with data flipping $(n_{\rm flip})$ and noise addition $(n_{\rm noise})$ are 2 and 8 folds more than the present values on the horizontal axis as mentioned above.

The basic trend here is that both mean $L_2$ error norm and standard deviation (colored surface) decrease as the number of training snapshots increases.
Although the cylinder wake data we used is governed by a periodic nature, large amount of training data is required for the sufficient accuracy.
Since we do not sample the original DNS data continuously for training data preparation, i.e., randomly extracted from 1000 snapshots with the time interval of $0.25$ dimensionless time, the snapshots in various cycles are contained in the training dataset as the number of original snapshots $n_{\rm DNS}$ increases.
Fukami et al.\cite{FFT2020} had observed in detail that even with the periodic data, the estimation accuracy is improved by feeding larger amount of training data since there may be a slight offset in phase among training data.
Especially through out bulking out techniques, the estimation ability is drastically improved at smaller number of snapshots --- the training data could be successfully augmented to generalize the neural network, as seen in figures \ref{fig_cd_result}$(b)$ and~$(c)$.

In contrast, for the larger number of training snapshots, it is striking that noise addition causes negative influence on the estimation accuracy.
The mean $L_2$ error norm, averaged among the three-fold cross validation, of the estimation through noise addition (red triangle plots) shows larger value comparing to other cases and also the range of standard deviation (red colored surface) is getting wider as the $n_{\rm noise}$ increases.
The finding here suggests that it is likely inappropriate to add the synthetic noise to  sufficient amount of training data since the neural network can already acquire the nature of dataset.

Similar observation can also be found in the use of data flipping.
The error of the estimation through data flipping (blue square plots) approximately converges to that of standard training (grey circle plots). This suggests that the amount of original training data for flipping is sufficient to learn the given input-output relationship.

\subsubsection{Super-resolution analysis}
\label{sec_cy-sr}

\begin{figure}
	\centering
	\includegraphics[width=\textwidth]{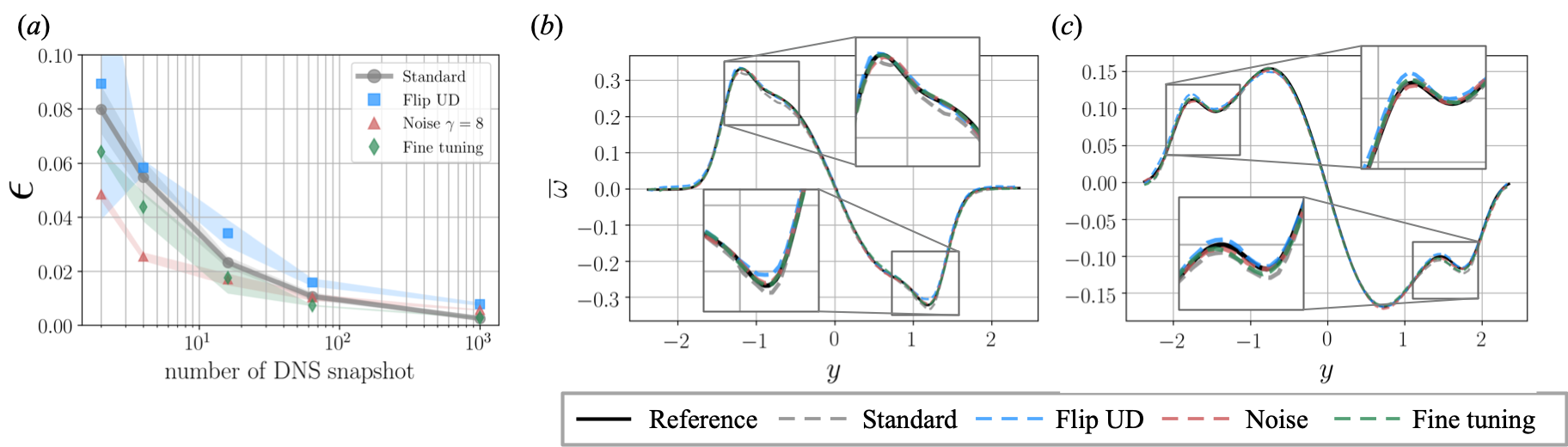}
	\caption{$(a)$ $L_2$ error dependence of super-resolution reconstruction for two-dimensional cylinder wake on the number of used DNS snapshots and data augmentation techniques. Mean vorticity profile of models trained with 4 training snapshots at $(b)$ $x=13.2$ and $(c)$ $x=15.7$.}
	\label{fig_sr_cylinder}
\end{figure}

To observe the behavior for high-dimensional output regressions, we then consider the super-resolution task ${\bm q}_{\rm HR}={\cal F}({\bm q}_{\rm LR})$ of two-dimensional cylinder wake.
Super-resolution analysis, originally developed in the field of image processing, aims to estimate high-resolution data from its low-resolution counterpart.
The idea had been applied to the fluid flow data by Fukami et al.~\cite{FFT2019a} in 2019.
The similar idea can also be applied to the universal closure modeling for LES and RANS \cite{maulik2017resolution,Duraisamy2021}.
Moreover, considering that the low-resolution data corresponds to sparse sensor measurements, it can be applied to a global field reconstruction task from the limited data~\cite{erichson2019,FukamiVoronoi}.
These applications also encourage us to obtain detailed information of the weather or ocean data~\cite{SGHK2021}, which is crucial for disaster prevention.
Here, the low-resolution data $\bm{q}_{\rm LR}$ are generated by average pooling operation for original DNS data $\bm{q}_{\rm HR}$ to become 1/8 resolution of the original data.

The $L_2$ error dependence on the number of training snapshots is presented in figure \ref{fig_sr_cylinder}$(a)$.
In this particular example, the data flipping (blue plots) has no clear advantage against the standard training process.
In contrast, the estimation can be improved by adding the Gaussian noise to the training data (red plots), especially at the smaller number of training snapshots.
The observation for the smaller $n_{\rm snapshot}$ is analogous to the $C_D$ estimation in figure~\ref{fig_cd_result}.
The trend can also be observed from the mean vorticity profile at certain $x$ positions, as shown in figures~\ref{fig_sr_cylinder}$(b)$ and $(c)$.
The results of standard process and data flipping slightly disagree with the reference vorticity, shown in black solid line, where the magnitude of vorticity is relatively large.
On the other hand, the results with noise addition and fine tuning show great agreement, as can be expected from figure~\ref{fig_sr_cylinder}$(a)$.

We also consider the application of transfer learning to this task.
The input and output data here are low-resolution (LR) and high-resolution (HR) data.
The training process, generally explained in section \ref{sec_tl}, for this particular super-resolution task can be expressed as,
\begin{eqnarray}
    &&{\bm w}_{\rm pre}={\rm argmin}_{{\bm w}_{\rm pre}}||{\bm q}_{\rm HR,local}-{\cal F}_{\rm pre}({\bm q}_{\rm LR,local};{\bm w}_{\rm pre})||_2 \nonumber\\
    &&{\bm w}_{\rm post}={\rm argmin}_{{\bm w}_{\rm post}}||{\bm q}_{\rm HR,global}-{\cal F}_{\rm post}({\bm q}_{\rm LR,global};{\bm w}_{\rm post})||_2 \nonumber,
\end{eqnarray}
where the initial weights for the posteriori model ${\bm w}_{\rm post,init}={\bm w}_{\rm pre}$, ${\bm q}_{\rm HR}$ and ${\bm q}_{\rm LR}$ are high-resolution and low-resolution data.

\begin{figure}
	\centering
	\includegraphics[width=0.95\textwidth]{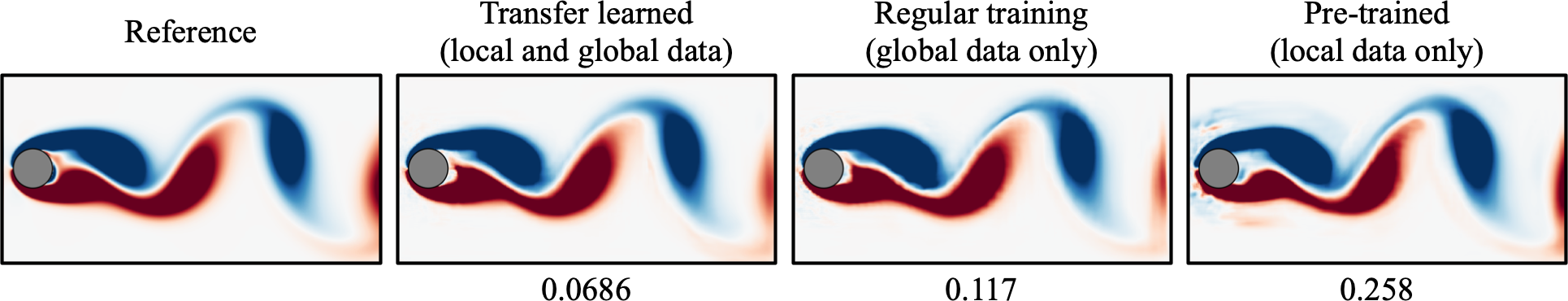}
	\caption{Super-resolved field estimated by the transfer learned model, the regular CNN and the pre-trained model. The values underneath each figure indicate the normalized $L_2$ error norm to the reference.}
	\label{fig_cy_sr}
\end{figure}

The super-resolved fields estimated by the transfer learned model ${\cal F}_{\rm post}$, the regular CNN, and the pre-trained model ${\cal F}_{\rm pre}$ are summarized in figure \ref{fig_cy_sr}.
All models are trained at $n_{\rm DNS}=2$.
Compared to the regular CNN trained with only global data, the transfer learned model reports approximately $5\%$ lower error.
This result demonstrates the strength of the transfer learning for fluid flow regression.
In addition, notable point here is that the pre-trained model ${\cal F}_{\rm pre}$ can reconstruct the whole field in reasonable accuracy despite that the model was trained with only local sub-domains.
This is because the CNN-based super-resolution reconstruction is scale invariant thanks to filter sharing over a whole domain in images.
It implies that the locally trained model ${\cal F}_{\rm pre}$ has acquired the generalized function of super-resolution over the spatial domain.
This feature of CNN may be utilized for the situation where the local data can only be handled due to users' CPU limitation.

\begin{figure}
	\centering
	\includegraphics[width=0.95\textwidth]{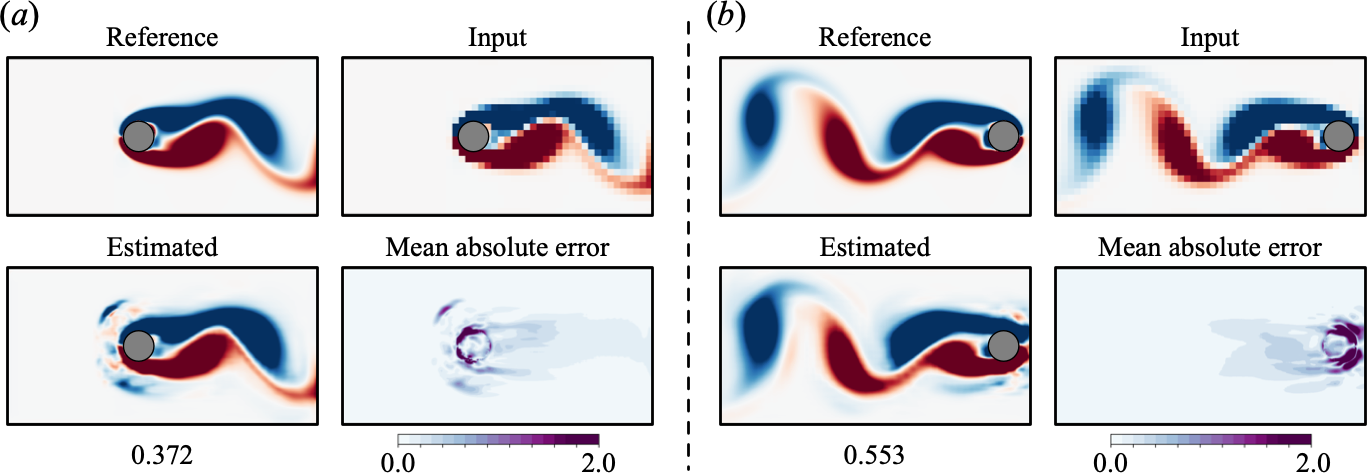}
	\caption{Application of trained machine learning model to unlearned alignment of bluff body for super-resolution analysis with a two-dimensional cylinder flow. $(a)$ downstream and $(b)$ inverse flow.}
	\label{fig_sr_tl_align}
\end{figure}

For further investigation, we test the model ${\cal F}_{\rm pre}$ to a flow with different alignment of the cylinder as shown in figure \ref{fig_sr_tl_align}, i.e., cylinder at $(a)$ downstream and $(b)$ inverse flow.
Both test data are prepared from the original DNS data.
As it can been seen, the model ${\cal F}_{\rm pre}$ successfully super resolves the wake region while the error concentrates on region around the body.
From these observations, we find that the model trained with local sub-domains had acquired the general function of super-resolution and it is invariant not only to the scale, but also to the different alignment of the cylinder.

\subsubsection{Temporal prediction}

The applicability of the present data augmentation techniques to machine learning-based temporal prediction is further investigated considering a cylinder wake and the NOAA sea surface temperature data.
Neural network-based surrogate modeling for a numerical simulation is one of the promising uses in nonlinear dynamical systems.
Thanks to its rapid estimation, a neural network-based method can estimate the future state of the dynamics with significantly shorter computational time compared to the traditional numerical approach.
For instance, Fukami et al.~\cite{FNKF2019} proposed an inflow driver for numerical simulation of turbulence using neural networks.
Another approaches consisted with long short-term memory~\cite{SGASV2019,HFMF2020b,HFMF2020a,nakamura2020extension} and sparse identification of nonlinear dynamics~\cite{bruntonsindy,fukami2020sparse} are also promising techniques in estimating the temporal evolution of the flows.

For the cylinder wake example, we train a machine learning model to estimate the field of next time step $t={(n+1)}\Delta t$ from the current state $t={n\Delta t}$, where the interval $\Delta t$ is 0.25 dimensionless time,
\begin{equation}
    {\bm q}^{(n+1)\Delta t}={\cal F}({\bm q}^{n\Delta t}).
\end{equation}
As the techniques for training data augmentation, we consider the all three techniques introduced in section \ref{sec:method}.
Note again that the amount of data is twice with the data flipping and it is eight times more with noise addition, similarly to the previous section.
For the transfer learning, the training process can be written as,
\begin{eqnarray}
    &&{\bm w}_{\rm pre}={\rm argmin}_{{\bm w}_{\rm pre}}||{\bm q}^{(n+1)\Delta t}_{\rm local}-{\cal F}_{\rm pre}({\bm q}^{n\Delta t}_{\rm local};{\bm w}_{\rm pre})||_2 \nonumber\\
    &&{\bm w}_{\rm post}={\rm argmin}_{{\bm w}_{\rm post}}||{\bm q}^{(n+1)\Delta t}_{\rm global}-{\cal F}_{\rm post}({\bm q}^{n\Delta t}_{\rm global};{\bm w}_{\rm post})||_2,
\end{eqnarray}
where ${\bm w}_{\rm post,init}={\bm w}_{\rm pre}$.

\begin{figure}
	\centering
	\includegraphics[width=\textwidth]{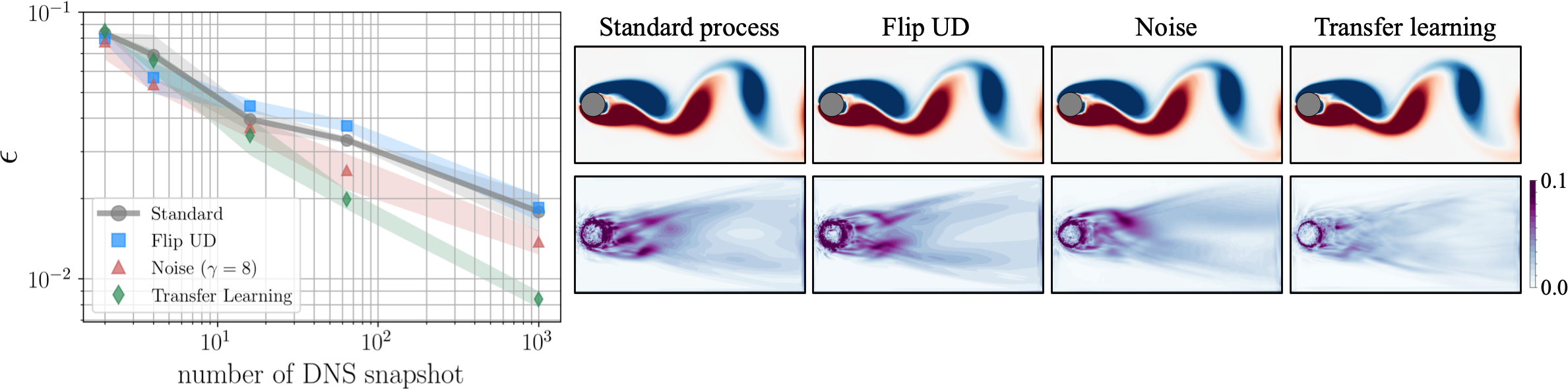}
	\caption{$L_2$ error dependence of temporal prediction for two-dimensional cylinder wake on the number of used DNS snapshots and data augmentation techniques. Representative fields with time-ensemble averaged root mean squared error at $n_{\rm DNS}=64$ are shown in the right.}
	\label{fig_cy_time}
\end{figure}
\begin{figure}
	\centering
	\includegraphics[width=\textwidth]{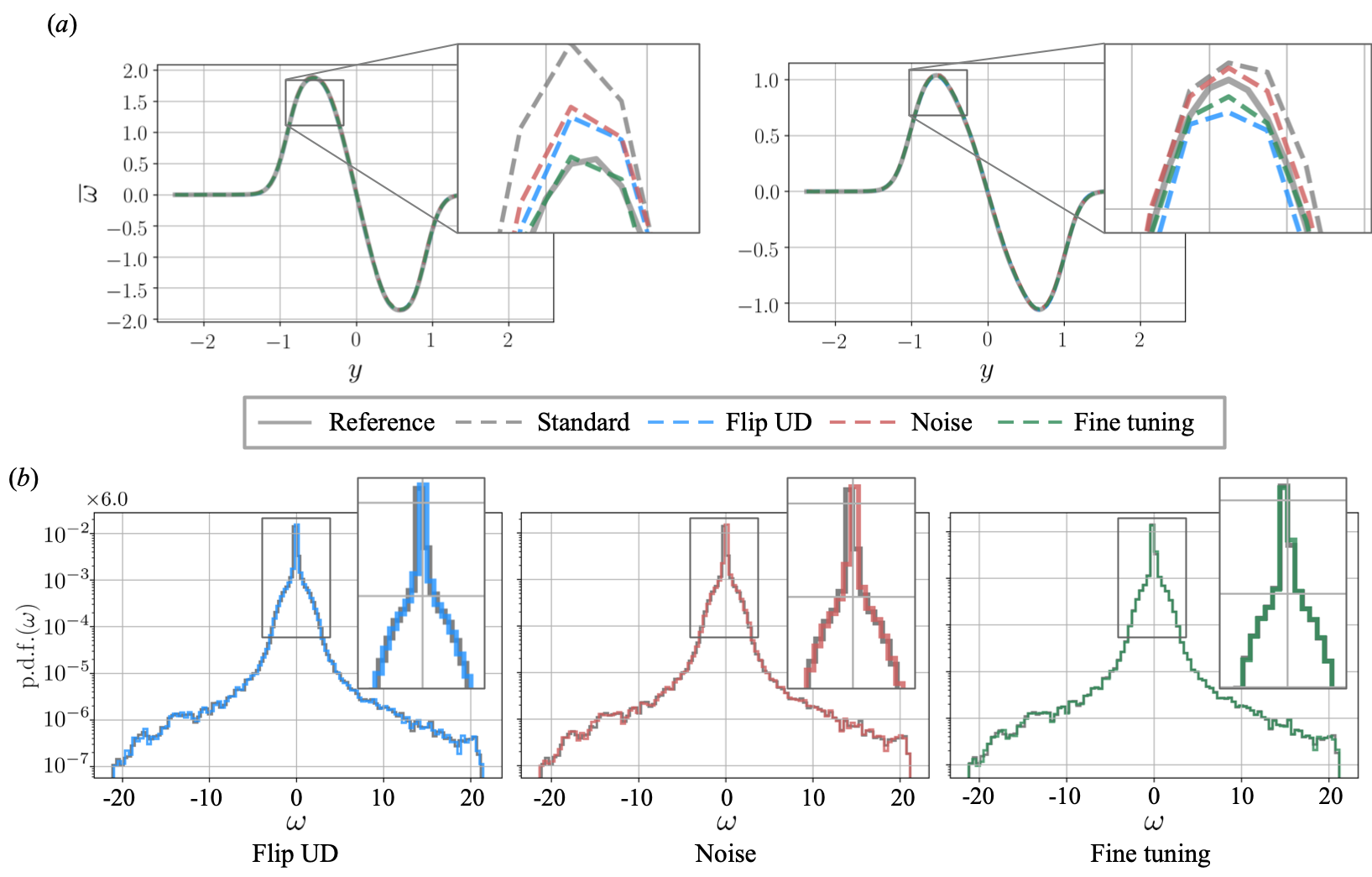}
	\caption{Results of the temporal prediction model with 64 training snapshots of cylinder wake. $(a)$ Mean vorticity profile at $x=10.7$ and $11.5$ and $(b)$ Probability density function of the vorticity field for each method.}
	\label{fig_cy_time_omg_pdf}
\end{figure}

The error plot of all cases are shown in figure \ref{fig_cy_time}.
Analogous to the previous problem settings, i.e., $C_D$ estimation in figure \ref{fig_cd_result} and super-resolution in figure \ref{fig_sr_cylinder}, the basic trend shows that mean $L_2$ error norm and standard deviation, among the three-fold cross validation, decrease as the number of training snapshot increases.
As for the data flipping, it is observed that it shows no clear improvement against the standard process except for $n_{\rm DNS}=4$.
In contrast, the lower errors are reported on every $n_{\rm DNS}$ by adding the Gaussian noise to the training data.
This trend is unique among the other problem settings, i.e., $C_D$ estimation (figure \ref{fig_cd_result}), super-resolution task (figure \ref{fig_sr_cylinder}) and temporal estimation for sea surface temperature data, which will be discussed later.
The common trend observed through these other problem settings is that the estimation ability is improved with smaller number of snapshots while no significant difference (or even worse results) would be reported with larger number of snapshots comparing to the standard procedure.
This variation of trends among these cases suggests that care should be taken whether noise addition would be suitable for their particular situations or not, by considering target flows, problem settings, number of original snapshots and etc.

Furthermore, the model trained through the transfer learning shows striking results.
For the smaller number of snapshots, the transfer learned model shows no significant advantage to the standard process, although the error becomes even lower than the result of noise addition for the larger number of snapshots.
Unlike the other techniques, the difference between the standard process and the transfer learning is only the training process.
The input/output realization for the posteriori model ${\cal F}_{\rm post}$ is the same as that of standard process.
In other words, the only difference is whether the initial weight was set randomly or obtained through pre-training with local data.
Analogous to the super-resolution task (section \ref{sec_cy-sr}), we find that the pre-training using locally divided data can be one of the considerable tools to augment the generalizability of neural networks for fluid flow regression.
Moreover, the model trained via transfer learning reports slightly narrower range of standard deviation compared to the result with noise addition, indicating the training process is more stable over the cross validations.
The detailed observations for each method are summarized in figure~\ref{fig_cy_time_omg_pdf}.
The clear advantage of the transfer learning can also be found from the comparison of estimated mean vorticity at $x=10.7$ and $x=11.5$ shown in figure~\ref{fig_cy_time_omg_pdf}$(a)$.
The zoomed-in figures exhibit that the model with transfer learning shows better agreement with the reference data compared to the other cases.
The probability density function, shown in figure~\ref{fig_cy_time_omg_pdf}$(b)$ of the vorticity field also shows the great capability of transfer learning.
While the distributions estimated with data flipping and noise addition slightly mismatch with the reference data, the model trained via transfer learning shows almost perfect estimations on high probability components.

\begin{figure}
	\centering
	\includegraphics[width=\textwidth]{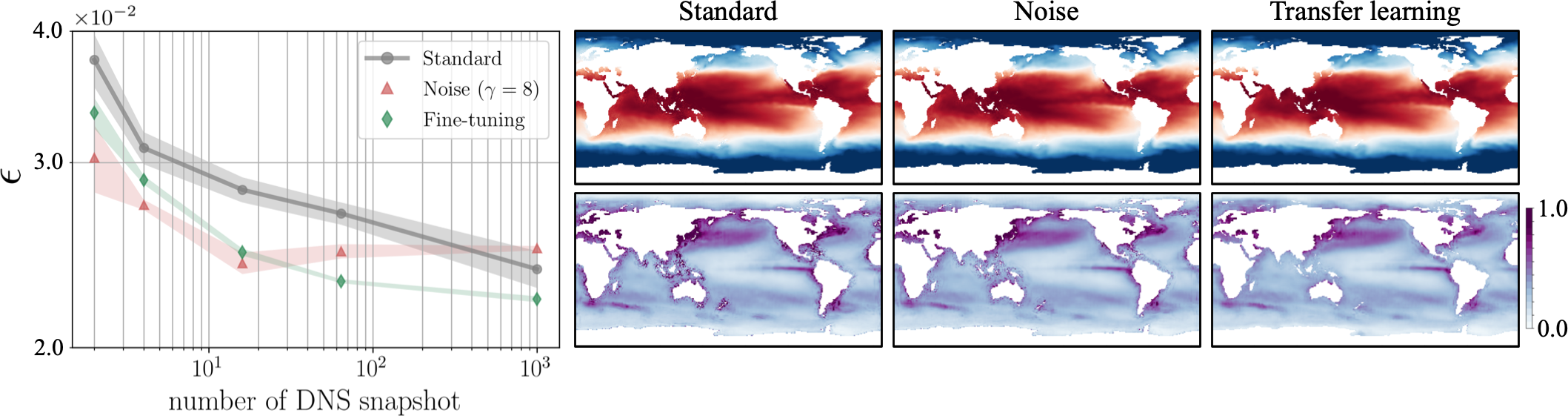}
	\caption{The dependence of the $L_2$ error on the number of used snapshots with the covered augmentation techniques of temporal prediction for sea surface temperature. Representative fields at $n_{\rm DNS}=64$ are shown in the right.}
	\label{fig_sst_time}
\end{figure}

Similar trends can also been seen with the result of temporal {prediction} of sea surface temperature data.
For the temporal prediction of sea surface temperature data, the model is trained to estimate the state of one week ahead thorough standard process, noise addition, and transfer learning.
Note that we do not use the data flipping since the geographical data here is asymmetric in both longitude and latitude axes.
Let us present in figure \ref{fig_sst_time} the dependence of the $L_2$ error on the number of used snapshots with the covered augmentation techniques.
With most cases, the error can be successfully decreased utilizing both noise addition and transfer learning.
Similar to the temporal prediction of the cylinder wake, the result with noise addition marks lower error for smaller number of snapshots, while the result of transfer learning becomes superior as the training snapshots increases.
The result of noise addition shows the similar trend to that of $C_D$ estimation on cylinder wake (figure \ref{fig_cd_result}).
Among $n_{\rm DNS}=16$ to $64$, the error becomes larger despite the increase of training snapshots.
As discussed in section \ref{sec:cdbulk}, this is likely a demarcation where the influence of synthetic noise turns to be negative as the amount of original training data becomes sufficient.

\begin{figure}
	\centering
	\includegraphics[width=0.75\textwidth]{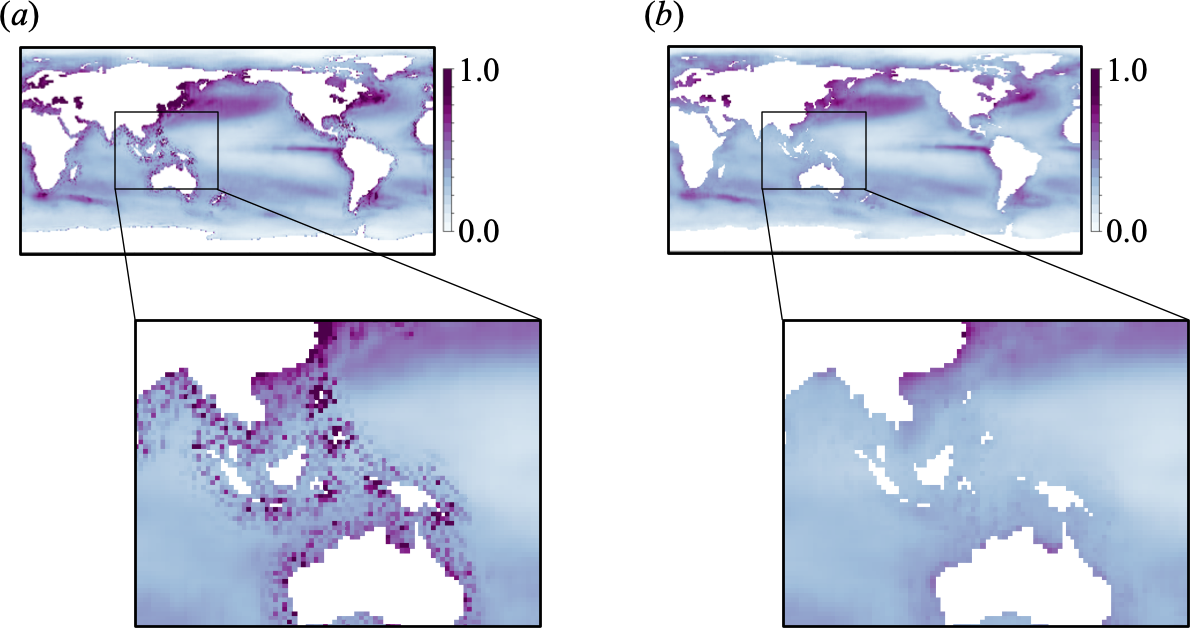}
	\caption{Local root mean squared error calculated from the results obtained by $n_{\rm ref}=1000$ through $(a)$ standard process and $(b)$ transfer learning for the temporal prediction of sea surface temperature.}
	\label{fig_sst_tl}
\end{figure}
\begin{figure}
	\centering
	\includegraphics[width=\textwidth]{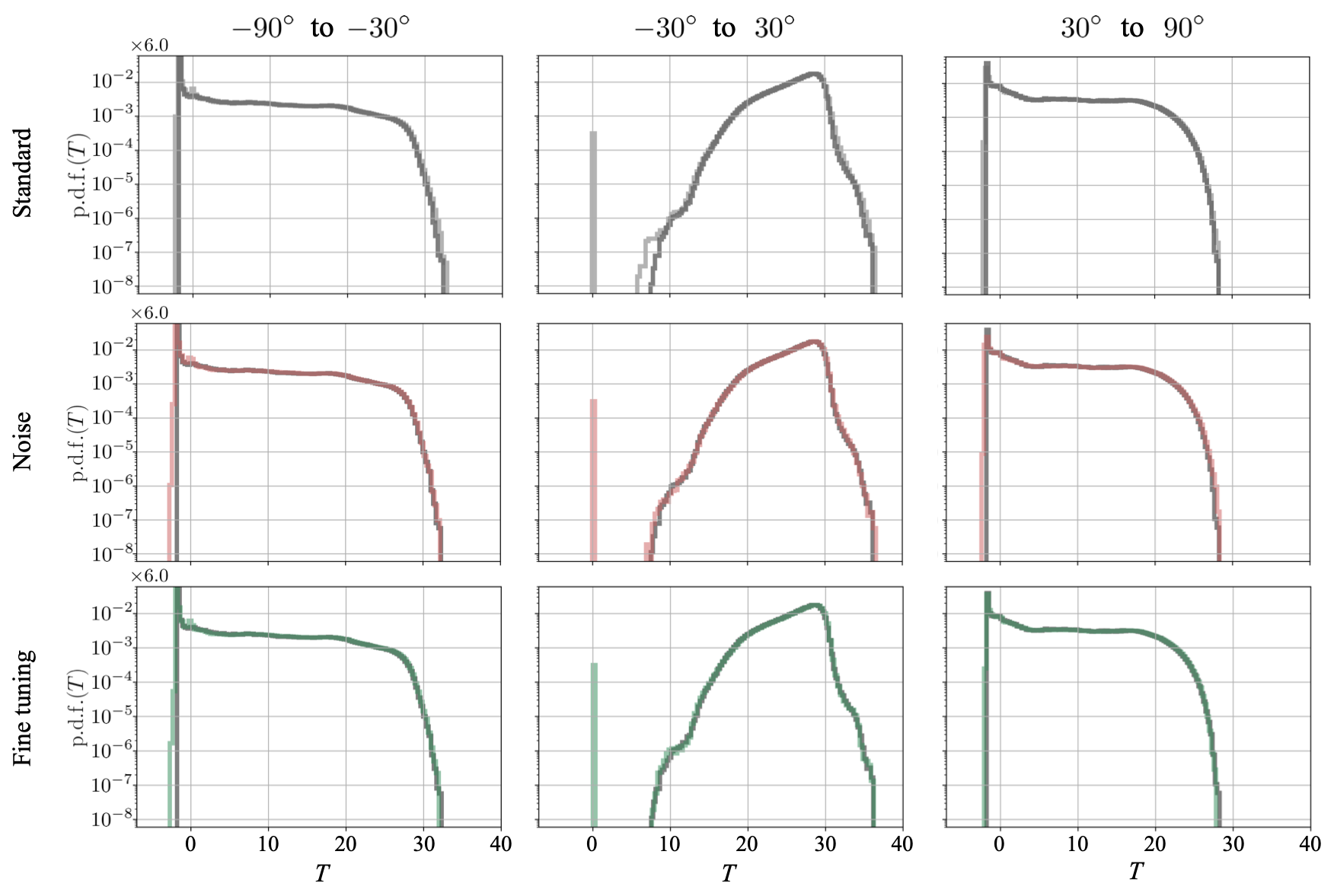}
	\caption{Probability density function (p.d.f.) of the temperature field. The p.d.f. is shown individually in three different latitude band: 1) $-90^\circ$ to $-30^\circ$, 2) $-30^\circ$ to $30^\circ$, and 3) $30^\circ$ to $90^\circ$. The reference data is shown in dark grey for comparison.}
	\label{fig_sst_pdf}
\end{figure}

On the other hand, the model trained through transfer learning shows notably lower error for all $n_{\rm DNS}$ cases.
Also, the standard deviation of $L_2$ error norm becomes significantly narrower comparing to not only the result of noise addition, but also the standard process.
What is striking here is that significant improvement can be observed on region around the continents.
As shown in figure \ref{fig_sst_tl}, the root mean squared error around the continents are smoothed via transfer learning compared to the standard training process, i.e., trained with global data only.
Since the transfer learned model ${\cal F}_{\rm post}$ is first trained with local data, the model might be able to acquire the better estimation ability for local manner, e.g., influence of the continents.

The improvement in the estimation with noise addition and fine tuning can also be found from the probability density function, as shown in figure~\ref{fig_sst_pdf}.
We here consider three different latitudinal band, i.e., 1. $-90^\circ$ to $-30^\circ$, 2. $-30^\circ$ to $30^\circ$, and 3. $30^\circ$ to $90^\circ$.
While the result of standard training process reports the apparent mismatch on low-probability components for the area between $-30^\circ$ and $30^\circ$, the models trained with noisy data and fine tuning show nice agreement with the reference distribution.

\subsubsection{Generalization for unlearned data}
\label{sec_two-cy}

\begin{figure}[t]
	\centering
	\includegraphics[width=0.95\textwidth]{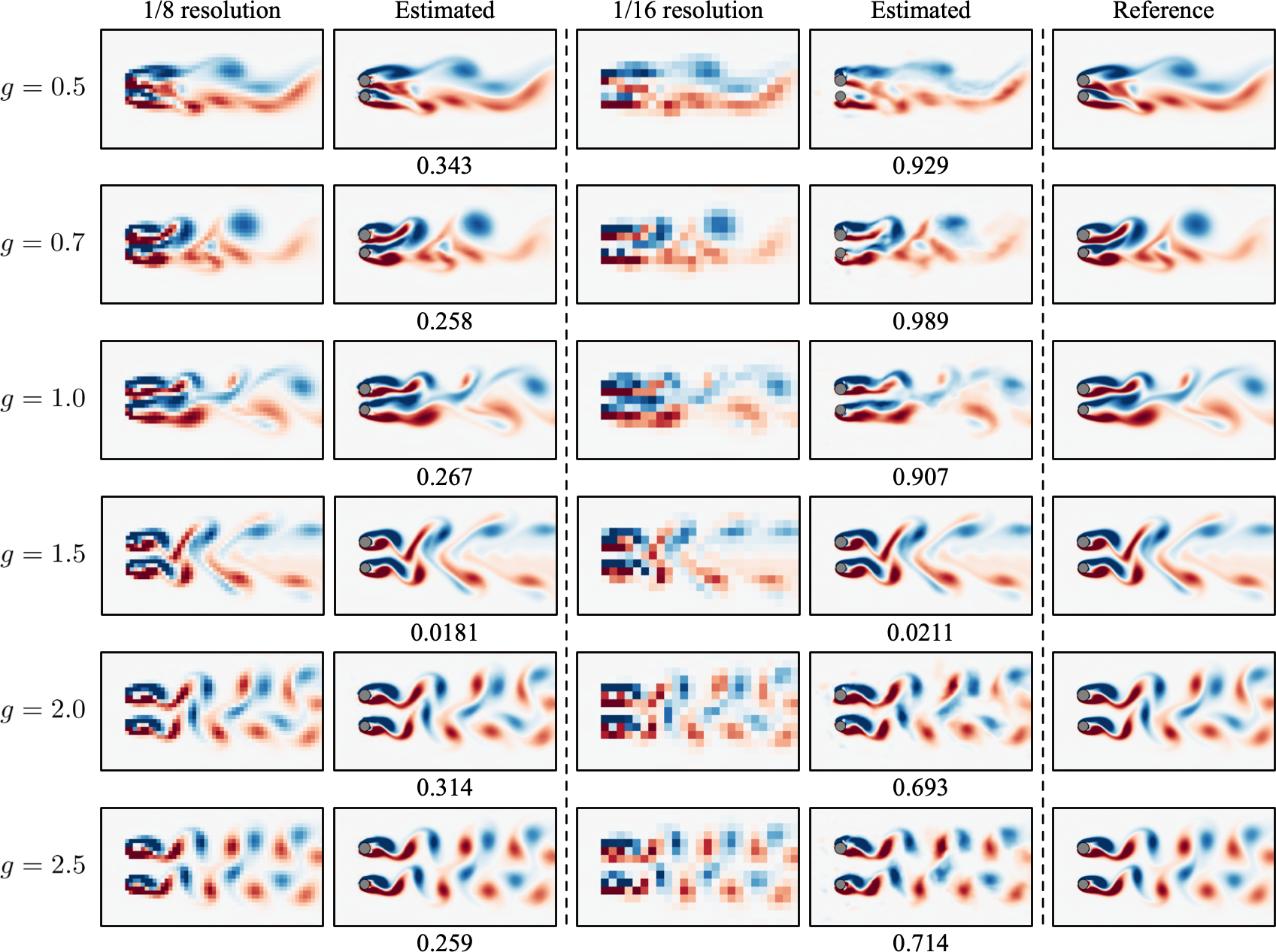}
	\caption{Super resolved fields of two-parallel cylinders estimated by the model trained with global data at $g=1.5$.}
	\label{fig_g2g_tr1.5}
\end{figure}

We demonstrated several data augmentation methods above while considering canonical flows, i.e., two-dimensional cylinder wake and sea surface temperature data.
In this section, let us investigate the applicability of machine learned models to unlearned state of the flow utilizing a two-parallel cylinder wake.
As presented in section \ref{subsection:2Psetup}, the flow over the two-parallel cylinders will change drastically by adjusting the gap ratio $g$.
Utilizing this unique characteristic of the flow, we here examine the generalizability of the neural network via super-resolution task with the up-sampling CNN introduced in figure \ref{fig_schm}$(c)$.
We consider two input coarseness, i.e., 1/8 and 1/16 resolution of original as shown in the first and the third rows of figure \ref{fig_g2g_tr1.5}.
The grid number for 1/8 and 1/16 resolution data are $30\times56$ and $15\times28$, respectively.
To observe the applicability of trained model to unlearned states of the flow, we construct three models whose training data include only single case in terms of $g$ for each model, i.e, (i) $g=1.5$, (ii) $g=0.5$, and (iii) $g=2.5$.
For instance, the model for case (i) is trained only using the flow regime at $g=1.5$ and tested covering all cases of $g$.

We first select the flow at $g=1.5$ for the training data as shown in figure \ref{fig_g2g_tr1.5}.
The estimation ability at $g=1.5$ is significantly better than the other test cases since the model is trained with the same $g$, although the test time range at $g=1.5$ is excluding the training process.
For the 1/8 resolution data, the trained network is able to successfully reconstruct the low-resolution data of all cases with the $L_2$ error rate of less than 0.35.
The representative super-resolved flow fields are also in excellent agreement with the reference data.
On the other hand, although the trained model is able to catch the rough trend of data in all test cases, the $L_2$ error norms for the 1/16 resolution case are relatively higher than that of 1/8 resolution cases, which implies the influence on the coarser input data.

\begin{figure}[t]
	\centering
	\includegraphics[width=0.8\textwidth]{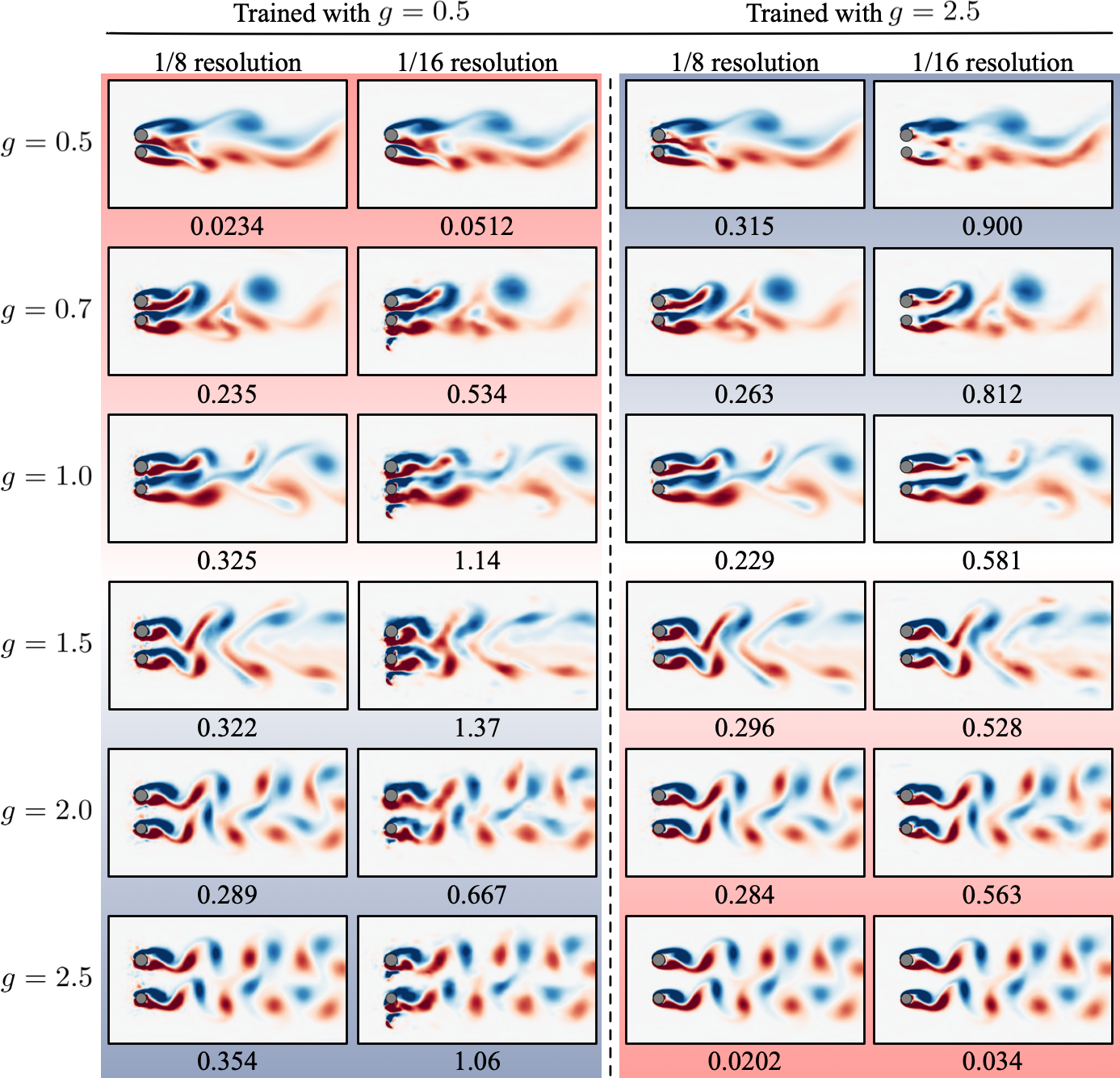}
	\caption{Super-resolved fields of two-parallel cylinders estimated by the model trained with global data at $g=0.5$ and $g=2.5$. Red background indicates flow fields that have similar $g$ to training data, while gray background indicates flow fields that are far from training range in terms of $g$.}
	\label{fig_g2g_tr0.5_2.5}
\end{figure}
\begin{figure}[t]
	\centering
	\includegraphics[width=0.9\textwidth]{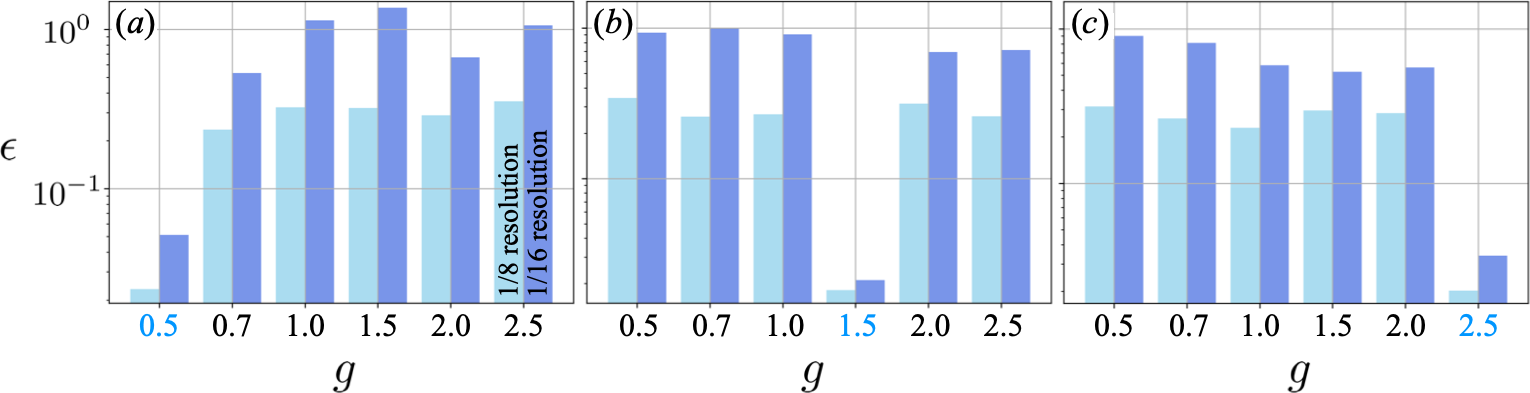}
	\caption{$L_2$ error norm of estimated fields by the model trained with global data at $(a)\ g=0.5, (b)\ g=1.5$ and $(c)\ g=2.5$. The factor $g$ used for training is colored by blue in the horizontal axes.}
	\label{fig_g2g_errorbar}
\end{figure}

The aforementioned trend --- the performance of the model is affected by the choice of training data --- can also be observed with the model trained at $g=0.5$ and $2.5$, as shown in figure \ref{fig_g2g_tr0.5_2.5}.
For both cases, the model can reconstruct the field from 1/8 resolution data with reasonable accuracy even with the flow field at unlearned $g$.
The estimation for the 1/16 resolution data is tougher than that for the 1/8 resolution, especially for the region around cylinders, which is likely because the location of each cylinder differs among fields at different $g$.
The observed sensitivity here for the location of bluff body is analogous to our finding of super-resolution analysis for a single cylinder in figure \ref{fig_sr_tl_align}.

\begin{figure}
	\centering
	\includegraphics[width=0.9\textwidth]{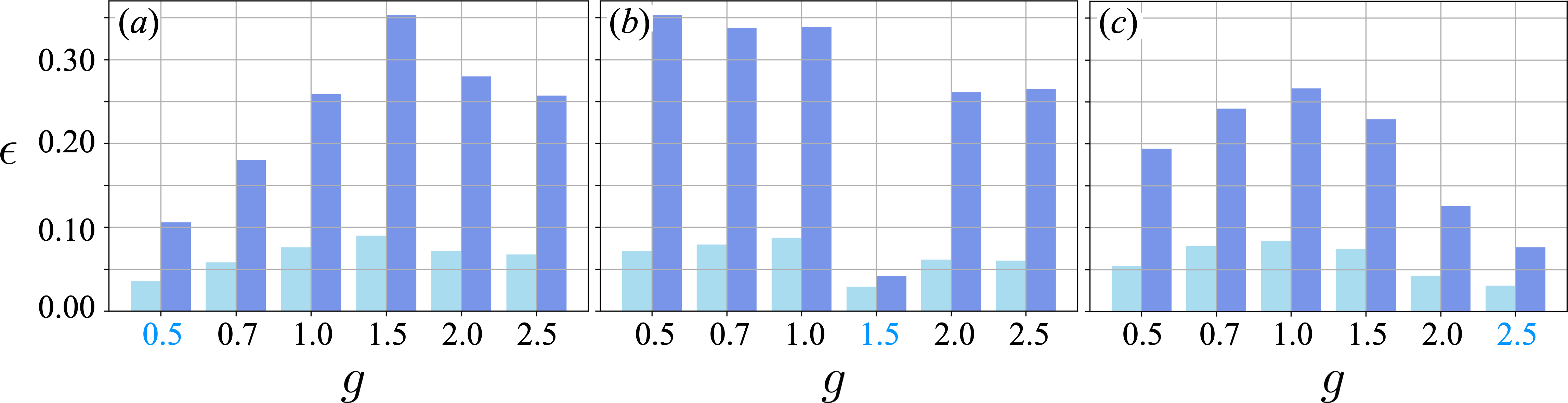}
	\caption{$L_2$ error norm calculated with wake region of estimated fields by the model trained with global data at $(a)~g=0.5, (b)~g=1.5$ and $(c)~g=2.5$. The factor used for training is colored by blue in the horizontal axes.}
	\label{fig_g2g_wake-region}
\end{figure}
\begin{figure}
	\centering
	\includegraphics[width=0.9\textwidth]{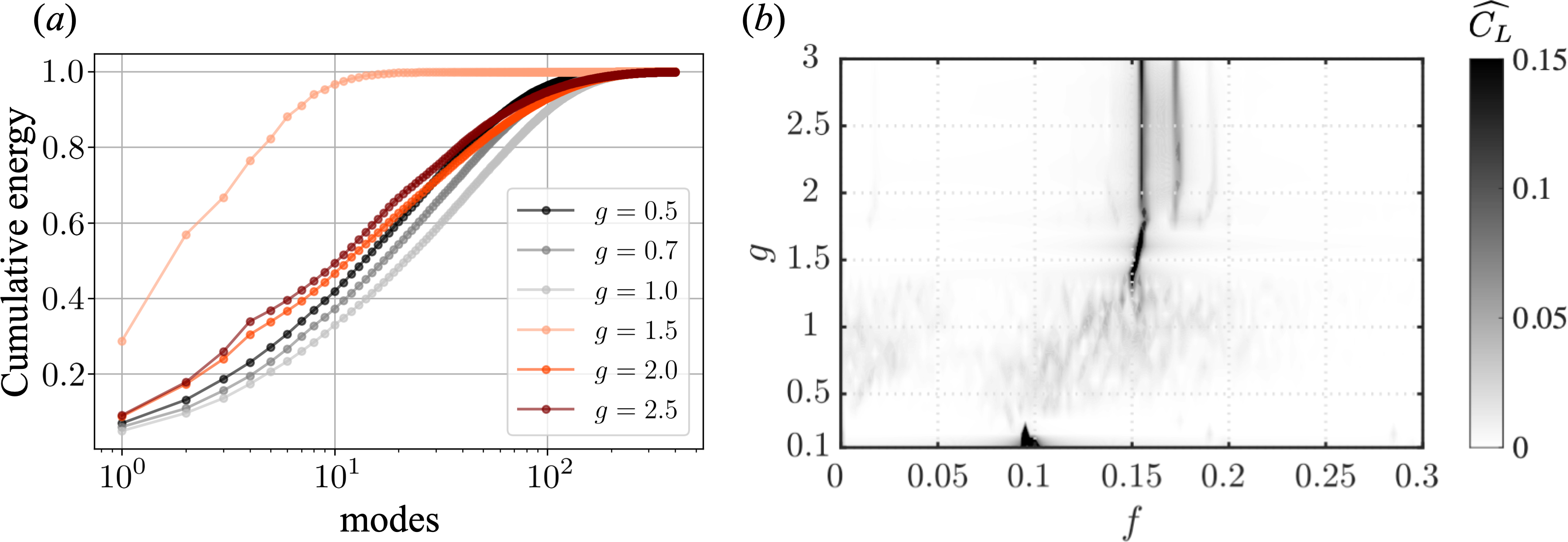}
	\caption{$(a)$ Singular value spectrum of two-parallel cylinders wake at each $g$. $(b)$ Amplitude spectrum of the lift coefficients. Here, $C_L=(C_{l,1}+rC_{l,2})/(1+r)$.}
	\label{fig_twocy_sv}
\end{figure}

To avoid the influence on the region around cylinders and focus on wake reconstruction, we also check the $L_2$ error norm on only wake region{, i.e., $4.38 \leq x \leq 19.0$}, as listed in figure \ref{fig_g2g_wake-region}.
For the model trained at $g=1.5$, the $L_2$ error on flows at unlearned $g$ is relatively high.
Even for flows with close gap ratio, i.e., $g=1.0$ and $2.0$, the error rate is approximately same as the other cases, which is a unique trend compared to the performance of the model trained at $g=0.5$ and $2.5$.
This is likely because the flow at $g=1.5$ is governed by synchronized vortex shedding between the two wakes. 
As demonstrated by the singular value spectrum in figure \ref{fig_twocy_sv}$(a)$, the energy convergence of the flow at $g=1.5$ is significantly faster than that of the other cases, meaning the flow can be represented with smaller number of spatial modes.
Since the flow at $g=1.5$ does not contain complex structures as the other cases, the estimation ability on those cases became lower, even with the cases at similar $g$.
In contrast, the models trained at $g=0.5$ and $2.5$ report smaller errors, especially on flows at similar $g$.
The similarity here can be found from the amplitude spectrum of the lift coefficients in figure \ref{fig_twocy_sv}$(b)$, which shows the chaotic nature of flow for $g=0.5$, 0.7 and 1, the synchronization at $g=1.5$, and the quasi-periodicity at $g=2.0$ and 2.5.
Summarizing above, care should be taken for properly choosing the training data considering various factors of target flows.

\section{Concluding remarks}
\label{sec:conclusion}

We demonstrated several techniques for encouraging the practical use of neural networks on fluid flow problems.
We first focused on visualization inside neural networks from the perspective of interpretability, considering two techniques, i.e., layer visualization and use of Grad-CAM.
The great ability of them could be appreciated from observing the grounds of the estimation.
Especially, the Grad-CAM offered us a clear insight of the crucial region for the estimation.
The use of various data augmentation techniques for training dataset, i.e., data flipping, noise addition and transfer learning, were also considered with canonical fluid flow regressions.
Among the covered techniques, we also found that transfer learning through local data can be a great candidate for improving the estimation ability drastically and stably in most of the cases.
Lastly, we investigated the generalizability of the neural networks for unlearned data through super-resolution analysis.
The trained network was able to catch the rough structure of the flow field even with the flows that are not included in the training data.
In particular, the lower error was reported on the flows which has similar characteristics to the training data according to a singular value spectrum.
Regarding the variation in data, we investigated the capability of the machine-learned model for unseen data by considering the temporal evolution of a flow around a bluff body with various different shapes in our previous paper~\cite{HFMF2020a}.
Moreover, the dependence of the model performance on Reynolds number is also investigated in~\cite{HFMF2020b}.
These observations also tell us that when the test situation is not too far from the training situation, and the training data are sufficiently given, a machine-learning model can hold the generalizability even for unseen data.

Since fluid flow phenomena contain highly nonlinear and chaotic nature, the techniques applied to the fluid flow data in the present study can be generalizable for the applications in a wide range of mechanical engineering.
In fact, a machine-learning-based temporal prediction technique proposed by our research group~\cite{HFMF2020b,HFMF2020a,nakamura2020extension} has recently been applied to the field of robotics to predict soil deformation in bucket excavation~\cite{ishigami2021}.
Such propagation of techniques from fluid mechanics motivates us that our proposed technique for highly nonlinear dynamics can be applied to a wide range of science.
Moreover, as demonstrated in the manuscript, the methods can be applied not only to fluid flows, but also for geophysical observation.
We can expect the applicability of our techniques to other types of geophysical data, e.g., weather and temperature field, as well.
We refer the enthusiastic readers to our previous paper which investigates the various parameter settings of convolutional neural network, which are utilized in all models covered in this study~\cite{MFZNF2021}.

Although we focused on the processing methods to reduce the amount of training data, other perspectives may also be considered to achieve the same goal.
For instance, a physics-informed neural network (PINN) \cite{raissi2020hidden} has recently been attracting attention since it can take a constraint based on physical laws as a loss function.
Thanks to this characteristic, it is highly expected that models with the concept of PINN can be trained accordingly from a small amount of training data while satisfying the physical laws.
Otherwise, the use of other sophisticated neural networks, e.g., graph neural network \cite{wu2020comprehensive} and reservoir computing \cite{inubushi2019transferring}, may be one of the possible candidates to improve the interpretability and generalizability, although careful choice is required depending on users' problem settings.
Moreover, from the perspective of data reconstruction from limited sensors, an optimal sensor placement derived with the theoretical manner can also drive a practical utilization of neural networks \cite{manohar2018data,NYNSN2020,saito2020data}.
We hope that our proposed techniques of internal procedure visualization and data bulking are able to encourage the practical use of neural networks in the fluid dynamics community, by combining with these aforementioned tools.

\begin{acknowledgements}
This work was supported from Japan Society for the Promotion of Science (KAKENHI grant number: 18H03758, 21H05007).
\end{acknowledgements}

%
\section*{Conflict of interest}

The authors declare that they have no conflict of interest.

\bibliographystyle{unsrt}
\bibliography{references}   


\end{document}